\documentclass{emulateapj}
\usepackage{graphicx}
\usepackage{epstopdf}
\usepackage{color}
\begin{document}
\shorttitle{HR 8799b}
\title{Near-Infrared Spectroscopy of the  Extrasolar Planet HR 8799 \lowercase{b}}
\author{Brendan P. Bowler,\altaffilmark{1} Michael C. Liu,\altaffilmark{1} Trent J. Dupuy,\altaffilmark{1} and Michael C. Cushing\altaffilmark{2}}
\email{bpbowler@ifa.hawaii.edu}

\altaffiltext{1}{Institute for Astronomy, University of Hawai`i; 2680 Woodlawn Drive, Honolulu, HI 96822, USA}
\altaffiltext{2}{Jet Propulsion Laboratory, California Institute of Technology; 4800 Oak Grove Dr., Mail Stop 264-723, Pasadena, CA 91109, USA}

\begin{abstract}
We present 2.12-2.23~$\mu$m high contrast integral field spectroscopy of the extrasolar planet HR~8799~b. Our observations were obtained with OSIRIS on the Keck~II telescope and sample the 2.2~$\mu$m CH$_4$ feature, which is useful for spectral classification and as a temperature diagnostic for ultracool objects.  The spectrum of HR 8799 b is relatively featureless, with little or no methane absorption, and does not exhibit the strong CH$_4$ seen in T dwarfs of similar absolute magnitudes. The spectrum is consistent with field objects from early-L to T4 (3~$\sigma$ confidence level), with a best-fitting type of T2. A similar analysis of the published 1-4 $\mu$m photometry shows the infrared SED matches L5-L8 field dwarfs, especially the reddest known objects which are believed to be young and/or very dusty. Overall, we find that HR~8799~b has a spectral type consistent with L5-T2, although its SED is atypical compared to most field objects.  We fit the 2.2~$\mu$m spectrum and the infrared SED using the Hubeny \& Burrows, Burrows et al., and Ames-Dusty model atmosphere grids, which incorporate nonequilibrium chemistry, non-solar metallicities, and clear and cloudy variants.  No models agree with all of the data, but those with intermediate clouds produce significantly better fits.  The largest discrepancy occurs in the $J$-band, which is highly suppressed in HR~8799~b.  Models with high eddy diffusion coefficients and high metallicities are somewhat preferred over those with equilibrium chemistry and solar metallicity.  The best-fitting effective temperatures range from 1300--1700~K with radii between $\sim$0.3--0.5~$R_\mathrm{Jup}$.  These values are inconsistent with evolutionary model-derived values of 800-900~K and 1.1--1.3~$R_\mathrm{Jup}$ based on the luminosity of HR~8799~b and the age of HR~8799, a discrepancy that probably results from imperfect atmospheric models or the limited range of physical parameters covered by the models.  The low temperature inferred from evolutionary models indicates that HR~8799~b is $\sim$400~K cooler than field L/T transition objects, providing further evidence that the L/T transition is gravity-dependent.  With an unusually dusty photosphere, an exceptionally low luminosity for its spectral type, and hints of extreme secondary physical parameters, HR~8799~b appears to be unlike any class of field brown dwarf currently known.

\end{abstract}

\section{Introduction}

High contrast adaptive optics (AO) observations on large telescopes are beginning to explore the outer architecture ($\gtrsim$10~AU) of planetary systems for the first time (e.g., \citealt{Biller:2007p19401}; \citealt{Lafreniere:2007p17991}; \citealt{Liu:2009p20202}; \citealt{Chauvin:2010p20082}).    The recent discoveries of giant planet systems around HR~8799 (\citealt{Marois:2008p18841}), Fomalhaut (\citealt{Kalas:2008p18842}), $\beta$~Pic (\citealt{Lagrange:2009p14794}; \citealt{Lagrange:2010p20870}), and 1RXS~J160929.1--210524 (\citealt{Lafreniere:2008p14057}; \citealt{Lafreniere:2010p20763}) by direct imaging have shown that planets exist at large separations ($\sim$10-330~AU) and are within reach of current telescope capabilities.  The advantages of direct imaging compared to the radial velocity technique are threefold: it is the only way to study planets on wide orbits; it allows for estimates of the true planet mass through luminosity measurements (rather than minimum masses as is the case for non-transiting radial velocity planets), and it enables follow-up photometric and spectroscopic observations to study extrasolar planetary atmospheres.

Multiwavelength observations of extrasolar planets can be used to infer their physical and chemical properties and test our understanding of planetary atmospheres through direct comparisons with theoretical atmospheric models.  Transmission spectroscopy and secondary eclipse observations of transiting exoplanets have led to the detections of several molecular species in the atmospheres of hot Jupiters, namely H$_2$O, CO$_2$, CO, and CH$_4$ (e.g., \citealt{Tinetti:2007p20414}; \citealt{Swain:2008p20415}; \citealt{Swain:2009p20416}).  These short-period transiting gas giants are also providing valuable tests of irradiated planetary atmospheric models and have shown evidence for temperature inversions (e.g., \citealt{Knutson:2009p18217}; \citealt{Spiegel:2009p18905}), global atmospheric dynamics (\citealt{Showman:2009p20418}), and non-equilibrium carbon chemistry (e.g., \citealt{Cooper:2006p20421}; \citealt{Swain:2010p20346}).   

At much larger separations of tens to hundreds of AU, observations of young ($\lesssim$100~Myr) substellar companions near and below the brown dwarf/planetary-mass limit ($\sim$13~$M_\mathrm{Jup}$) are revealing the physical properties of very low-mass objects through direct imaging and spectroscopy (e.g., AB~Pic~B: \citealt{Chauvin:2005p19642}, \citealt{Bonnefoy:2010p20602}; GQ~Lup~B: \citealt{Neuhauser:2005p18283}, \citealt{McElwain:2007p19412}, \citealt{Lavigne:2009p19386}; see Table 1 of \citealt{Zuckerman:2009p14792}).  For example, atmospheric model comparisons to near-infrared spectroscopy of the planetary-mass companion 2MASS~J1207-3932~b (\citealt{Chauvin:2004p19400}) is demonstrating the important role that dust may play in the atmospheres of young gas giants (\citealt{Mohanty:2007p6975}; \citealt{Patience:2010p20422}).

The system of three directly imaged giant planets orbiting HR 8799 presents an excellent opportunity to investigate the properties of extrasolar planetary atmospheres at young ages.  HR~8799 (=~HD 218396, V342~Peg) is a young\footnote{Recently, \citet{Moya:2010p20431} inferred an age of up to several Gyrs for HR~8799 using asteroseismology, which would imply that the HR~8799 companions are brown dwarfs ($\sim$30-40~$M_\mathrm{Jup}$ at 1~Gyr; \citealt{Baraffe:2003p588}).  However, the Frequency Ratio Method (\citealt{Moya:2005p20494}) used in that study is only accurate for a range rotational velocities (\citealt{Suarez:2005p20493}), which for HR~8799 limits the applicability to inclinations $\gtrsim$36$^{\circ}$ (\citealt{Moya:2010p20492}).  \citet{Reidemeister:2009p20163} infer an inclination between 20-30$^{\circ}$ based on a stability analysis of the companions and combining the range of true rotational velocities with $v$sin$i$.   Moreover,  \citet{Fabrycky:2010p20158} find that the orbiting companions must have masses $\lesssim$20~$M_\mathrm{Jup}$ based on a stability analysis.  Altogether this points to a younger age for the system.  Unfortunately, searches for wide stellar companions to HR~8799 which could further help constrain the age have not produced any promising candidates (\citealt{Close:2009p20168}; \citealt{Hinz:2010p20424}).  Additionally, the subsolar internal metallicity of HR~8799 inferred by \citet{Moya:2010p20492} may not be accurate because of the most probable inclination of the system.} (60$^{+100}_{-30}$~Myr; \citealt{Marois:2008p18841}), intermediate-mass (1.5~$\pm$~0.3 $M_{\odot}$; \citealt{Marois:2008p18841}), A-type\footnote{HR 8799 is a $\gamma$ Doradus variable star and belongs to the class of chemically peculiar $\lambda$ Bootis stars.  \citet{Gray:1999p20164} infer a spectral type of kA5~hF0~mA5 V $\lambda$ Boo based on detailed spectroscopic analysis, where ``k'', ``h'', and ``m'' refer to the Ca~II~K, hydrogen, and metallic line strengths and profiles, respectively.} star with an $Hipparcos$ parallactic distance of 39.4~$\pm$~1.1~pc (\citealt{vanLeeuwen:2007p12454}).  \citet{Marois:2008p18841} discovered three faint common-proper motion companions with projected separations of 68~AU (``b''), 38~AU (``c''), and 24~AU (``d'') and evolutionary model-derived masses of 7$^{+4}_{-2}$~$M_\mathrm{Jup}$, 10$^{+5}_{-3}$~$M_\mathrm{Jup}$, and 10$^{+5}_{-3}$~$M_\mathrm{Jup}$, inferred from their luminosities and the age estimate for HR~8799.  Subsequent analysis of previous imaging data from Keck II in 2007 (\citealt{Metchev:2009p19676}), Subaru in 2002 (\citealt{Fukagawa:2009p18543}), and the \emph{Hubble Space Telescope} in 1998 (\citealt{Lafreniere:2009p17982}) revealed several pre-discoveries of the HR~8799 planets and confirmed their common proper and orbital motions.  Dynamical modeling of this system suggests a small (but non-zero) inclination and two- or three-body resonances among the planets (\citealt{Gozdziewski:2009p19574}; \citealt{Reidemeister:2009p20163}; \citealt{Fabrycky:2010p20144}).  Thermal emission has been detected from a warm inner debris disk around HR~8799 at $\sim$10 AU and an outer massive disk at $\gtrsim$100 AU, implying an architecture analogous to our Solar System with giant planets surrounded by ``exo-asteroid'' and ``exo-Kuiper'' belts (\citealt{Sadakane:1986p20413}; \citealt{Zuckerman:2004p20412}; \citealt{Moor:2006p20096}; \citealt{Williams:2006p308}; \citealt{Rhee:2007p20409}; \citealt{Chen:2009p20407}; \citealt{Su:2009p20065}; \citealt{Reidemeister:2009p20163}).

The atmospheric properties of the HR 8799 planets have been investigated in several studies.   All three planets exhibit redder near-infrared colors than field T dwarfs with similar absolute magnitudes, which suggests the presence of dusty clouds in their atmospheres (\citealt{Marois:2008p18841}; \citealt{Metchev:2009p19676}; \citealt{Lafreniere:2009p17982}).  Additionally, non-equilibrium CO/CH$_4$ chemistry in the atmospheres of HR~8799~c and d has been suggested by \citet{Hinz:2010p20424} based on mid-infrared (3.3 $\mu$m, $L'$-band, and $M$-band) photometry.  Likewise, \citet{Janson:2010p20212} presented a low signal-to-noise (S/N) 3.88-4.10~$\mu$m spectrum of HR~8799~c and suggested that the disagreement with atmospheric models based on chemical equilibrium may hint at the influence of non-equilibrium chemistry.

Methane forms at temperatures $\lesssim$1400~K (e.g., \citealt{Fortney:2008p8729}), so under equilibrium conditions the spectroscopic signature of CH$_4$ provides an independent diagnostic of effective temperature for ultracool objects.  The presence of CH$_4$ absorption in the near-infrared is also the defining spectral signature of T dwarfs and therefore is a useful tool for spectral classification.  Medium-band photometry on and off the 1.6~$\mu$m methane band with the $CH_4s$ and $CH_4l$ filters shows no evidence for methane absorption in HR~8799~b and~c, although it may be present in HR~8799~d (\citealt{Marois:2008p18841}).  In this work we focus on the 2.2~$\mu$m absorption band of HR~8799~b, which is also sensitive to non-equilibrium CO/CH$_4$ chemistry at low temperatures and low surface gravities (\citealt{Hubeny:2007p14693}; \citealt{Fortney:2008p8729}).

We describe our adaptive optics (AO) observations with OSIRIS at Keck II and our data reduction procedure in $\S$\ref{sec:obs} and $\S$\ref{sec:datared}.  In $\S$\ref{sec:bdcomp} we compare field L and T dwarfs to our spectrum and previously published photometry for HR~8799~b, and we do the same with atmospheric models in $\S$\ref{sec:atmmod}.  Finally, we discuss our results in $\S$\ref{sec:discussion} and we provide a summary in $\S$\ref{sec:summary}.

\section{Observations}\label{sec:obs}

High contrast AO imaging observations are limited by quasi-static speckle noise at small angular separations ($\lesssim$1-2$''$), which results from slight imperfections in telescope optics and AO wavefront correction (\citealt{Racine:1999p19415}; \citealt{Marois:2000p19404}; \citealt{Marois:2008p17990}; \citealt{Oppenheimer:2009p18850}).  These speckle patterns pose many problems for directly imaging faint targets near bright stars.  Speckle noise is correlated and adds coherently, so longer exposure times do not improve contrast performance in this regime.  The speckles themselves can resemble real astronomical objects, making it difficult or impossible to distinguish between noise and true signal in a dataset.  Additionally, speckles are time-dependent phenomena spanning a range of durations (e.g., \citealt{Macintosh:2005p20425}; \citealt{Hinkley:2007p20367}).  Because of the temporal evolution of speckle patterns, no single observation of the point spread function (PSF) at a given time can accurately represent the PSF for observations taken over a long period of time.

As a result of these difficulties, there has been a concerted effort to develop observing and reduction techniques to overcome the limitations imposed by speckle noise.  One version of the spectral differential imaging technique (SDI; \citealt{Sparks:2002p19396}; \citealt{Thatte:2007p20429}) takes advantage of the wavelength-dependent nature of speckle positions to distinguish speckles from \emph{bona fide} objects, which remain stationary across all wavelengths.\footnote{This version of SDI makes the fewest assumptions about the planet's spectrum.  An earlier, simpler version of SDI (\citealt{Smith:1987p19403}; \citealt{Racine:1999p19415}; \citealt{Marois:2000p19404}; \citealt{Biller:2007p19401}) involves simultaneously imaging in two neighboring wavebands over which plant-star contrasts significantly vary, e.g. at the 1.6 $\mu$m CH$_4$ absorption band present in low-temperature objects.}  In principle, integral field spectrograph (IFS) observations can naturally utilize SDI (\citealt{Sparks:2002p19396}; \citealt{Berton:2006p19411}; \citealt{Thatte:2007p20429}; \citealt{Antichi:2009p19414}).  Yet to preserve the flux from real objects, SDI requires a fractional change in wavelength that is large enough to avoid self subtraction after de-magnification.  SDI with IFSs is therefore more practical for observations with broad wavelength coverage (i.e., broad-band filters rather than narrow-band filters).  Here we make use of a variant of the angular differential imaging method (ADI; \citealt{Liu:2004p17588}; \citealt{Marois:2006p18009}), which normally relies on rotation of the field of view when the telescope rotator is turned off.  This speckle subtraction technique makes it possible to distinguish between the speckle pattern, which remains stationary, and real objects, which rotate.  Our implementation of ADI is different, however, as we leave the rotator on to ensure that HR~8799~b remains in the same position on our detector (to avoid the planet rotating out of the narrow field of view of our observations; see below).  In this version of ADI the planet remains stationary while the speckles rotate.

\begin{figure}
  \resizebox{3.5in}{!}{\includegraphics{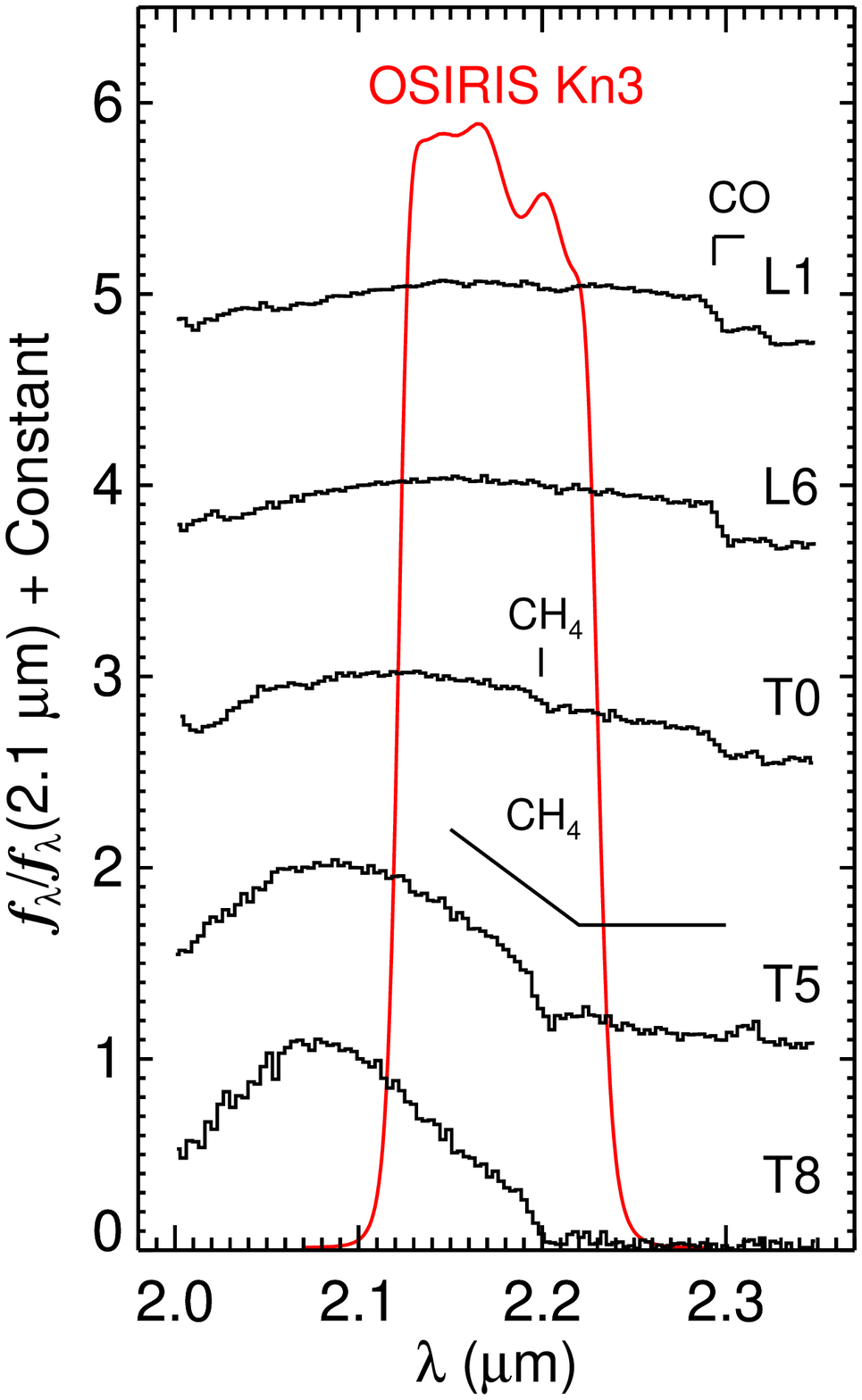}}
  \caption{Keck~II/OSIRIS Kn3 filter transmission profile.  Overplotted are IRTF SpeX/prism spectra of L and T dwarfs showing the increasing strength of the 2.2~$\mu$m CH$_4$ feature at cooler temperatures.    The spectra were obtained from the SpeX Prism Spectral Library and were originally published by \citet[2MASS J02271036--1624479: L1]{Burgasser:2008p14471}, \citet[ 2MASS J10101480--0406499: L6]{Reid:2006p19467}, \citet[SDSS J120747.17+024424.8: T0]{Looper:2007p819}, and \citet[2MASS J15031961+2525196: T5, 2MASSI J0415195-093506: T8]{Burgasser:2004p574}.  All data are normalized to 2.1~$\mu$m and offset by a constant. \label{fig:filtplot} } 
\end{figure}

Our observations of HR~8799~b were carried out with the OH-Suppressing InfraRed Integral-field Spectrograph (OSIRIS; \citealt{Larkin:2006p5567}) with natural guide star AO at Keck II on 2009 July~21~UT.  The weather was clear with a seeing estimate of $\sim$0\farcs7 in the optical.  OSIRIS IFS observations result in data cubes with spectral information over a large number (between $\sim$1-3$\times$10$^3$) of spatial positions.  The longer-lived  portion of the speckle pattern in the OSIRIS data cubes can be used to register the individual images in an ADI sequence.   Ideally we would obtain broad wavelength coverage over a large field of view  to maximize the number of  speckles in our observations, but there is a tradeoff between spectral coverage and field of view with IFS instruments as a result of limited space on the detector.  We chose the Kn3 filter (2.121-2.229~$\mu$m\footnote{These wavelengths correspond to the half-power points for the Kn3 filter.  The OSIRIS pipeline excludes wavelengths outside this region in the reduction.}; Figure \ref{fig:filtplot}) with the smallest plate scale (0$\farcs$02) to target the 2.2~$\mu$m CH$_4$ band with a moderate field of view ($\sim$0$\farcs$96$\times$1$\farcs$28) while simultaneously well-sampling the PSF.  Our filter choice was also based on AO performance, which is best in the $K$-band, and the fact that better CH$_4$ line lists exist at 2.2~$\mu$m than at 1.6~$\mu$m, thereby enabling more accurate constraints on physical parameters using atmospheric models.  The spectral resolving power ($R$ $\equiv$ $\lambda$/$\Delta\lambda$) of the OSIRIS data varies as a function of lenslet geometry (which depends on the filter), wavelength, and spatial position.  At the location of HR~8799~b in our observations, we estimate $R$~$\sim$~4300.\footnote{Based on data in the OSIRIS User's Manual, v.2.3 (http://irlab.astro.ucla.edu/osiris).}

We obtained 18 consecutive observations of HR~8799~b with exposure times of 300~s each for a total integration time of 90~min.   The airmass varied from 1.16 to 1.00 throughout our observations, spanning 13$^{\circ}$ of the rotation of the PSF in our field of view.  Immediately afterwards we obtained observations of the A0V standard star HD~208108 for telluric correction.  We took sky frames before and after the science and standard star sequence. The FWHM of the standard star is 2.4 spaxels at 2.2~$\mu$m, or 0$\farcs$048 with the 0$\farcs$02 plate scale (i.e., diffraction-limited).

\section{Data Reduction}\label{sec:datared}

\subsection{LOCI Processing}

\begin{figure}
  \vskip -.5 in
   \hskip -.5 in
  \resizebox{4.8in}{!}{\includegraphics{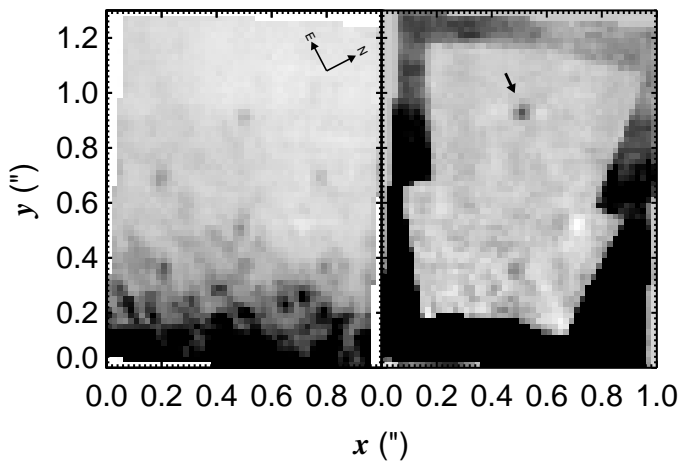}}
  \caption{\emph{Left}: Example of a median-combined OSIRIS data cube from a single 5 minute observation of HR~8799~b.  The planet is mostly indistinguishable from the speckle noise in this image.  Consecutive images show the speckles rotating but the planet fixed in the same location.  The location of HR~8799~A is $\sim$0.8$''$ below the edge of the array.  \emph{Right}: Fully processed image of HR~8799~b.  The planet (marked with an arrow) is clearly visible after the image reduction using LOCI.  This image is the median-combination of 18 individual observations for a total integration time of 90 minutes.  LOCI has been applied in two annular subsections using the parameters from $\S$\ref{sec:datared}.  The images are displayed with an asinh intensity stretch (\citealt{Lupton:2004p20516}).  Note that the object $\sim$0$\farcs$6 below the planet is an artifact from poor speckle subtraction.    \label{fig:adifig} } 
\end{figure}

Initial data reduction was performed using the OSIRIS data reduction pipeline (v.2.2; \citealt{Krabbe:2004p13521}). The OSIRIS  pipeline corrects for detector artifacts, performs sky subtraction and flat fielding, wavelength-calibrates the data, and assembles the 2D spectra from each lenslet into 3D data cubes.  Every resulting data cube contains 3063 spectra, each with 433 spectral channels and a spatial geometry of $\sim$48$\times$64 ``spaxels,'' or spatial pixels.  An example of a median-combined data cube from our observations is shown in Figure \ref{fig:adifig} (left panel).

The limited field of view of our OSIRIS configuration made registering the individual data cubes more difficult than for standard high contrast imaging observations because the OSIRIS field of view encompassed only a portion of the stellar PSF and did not include the peak. (HR~8799 was placed $\sim$0$\farcs$8 below the edge of the array.)   Slight drifting caused the image positions to move on a 1-2 spaxel scale over the course of our observations.  To register the data cubes we chose an observation near the middle of the ADI sequence and then used the relative rotation angle of the remaining observations to individually derive the $x$ and $y$ pivot coordinates about which the telescope rotated.  The rotation angle between exposures was computed from the FITS header information and remained a fixed parameter in our analysis.  We binned each data cube into 10 spectral channels and solved for a single pivot point position for each observation.  This was done in an iterative fashion using the \texttt{AMOEBA} algorithm (\citealt{Press:2007p13558}) by minimizing the rms of each de-rotated residual data cube (i.e., the fixed cube in the middle of our ADI sequence minus the cube being considered).  The rms was computed in an overlapping rectangular region of the residual data cube and included several bright speckles.  A visual inspection of each de-rotated data set confirmed the proper alignment of bright speckles.\footnote{Although the speckles were well aligned using our registration technique, there was a large variation in the inferred pivot point positions (usually differing by 5-20 spaxels, but even larger for a few of the exposures).  The reason for this discrepancy is that the pivot point position is highly sensitive to small adjustments when aligning two exposures using this technique.  The impact on the extracted spectrum was minor when the pivot point and rotation angle were fixed in the reduction; the main effect was to increase the spectral measurement errors, but the shape of the spectrum was preserved.}  We tested the registration with several choices for the number of spectral bins and found that 10 bins adequately compensated between S/N and number of spectral channels.

We performed speckle subtraction on our de-rotated data cubes using the ``locally optimized combination of images'' (LOCI) algorithm (\citealt{Lafreniere:2007p17998}; Figure \ref{fig:adifig}, right panel).  For each image, LOCI solves for the optimal linear combination of all the other images to build a reference frame for PSF subtraction. We median-combined each data cube into 10 spectrally binned frames and independently treated each binned spectral channel as images in an ADI sequence. Following \citet{Lafreniere:2007p17998}, we experimented with the different LOCI optimization parameters that control the shape and size of the optimization annular subsection ($g$ and $N_A$), the size of the subtraction subsection ($dr$), and the minimum displacement distance from the science target to avoid self-subtraction ($N_\delta$).  We systematically varied each LOCI parameter and the number of bins to study the influence on the resulting spectrum.  Adjusting the values of $N_A$ (from 100 to 400), $g$ (from 0.5 to 2.0), and $dr$ (from 2 to 10) had little effect on the spectrum.  Values of $N_\delta$ below 1.0 (in units of PSF FWHM) resulted in a smaller signal from the planet due to increased self-subtraction, but values of $N_\delta$ between 1.0-2.0 had only a minor impact.  Increasing the number of spectral bins resulted in a noisier spectrum, but the shape and amplitude of the spectrum remained unchanged.  We used $N_A$=$300, g$=1.0, $dr$=10, and $N_\delta$=1.0 for our final spectrum. 

After speckle subtraction, we performed aperture photometry on each individual data cube to extract the spectra of HR 8799 b.   Although the sky background should be zero as a result of using LOCI, we tested aperture photometry with and without sky subtraction as a precaution.  ADI+LOCI produces a slight negative trough on either side of the planet spectrum as a result of de-rotation and subtraction.  These troughs may affect sky subtraction using sky values computed from an annulus so we also tried sky subtraction using a sky value computed from a nearby box; the shape and amplitude of the extracted spectrum were preserved in all cases and the final flux measurements were all in agreement.  We chose an aperture radius of 2 spaxels with no sky subtraction for our final extracted spectrum.  The final spectrum was assembled by scaling each of the 18 spectra to the median-combined spectrum and then computing the median of the scaled spectra.

\begin{figure}
  \resizebox{3.5in}{!}{\includegraphics{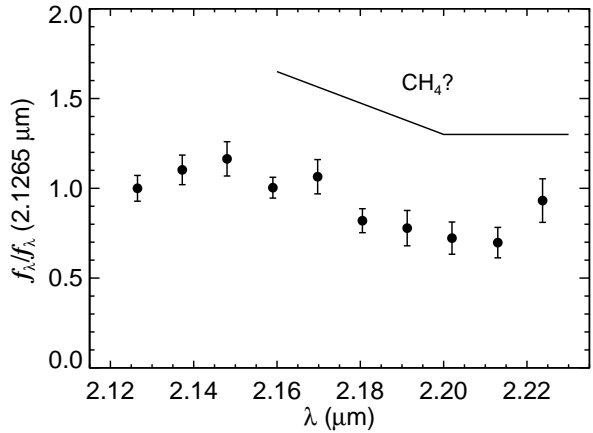}}
  \caption{Telluric-corrected OSIRIS spectrum of HR~8799~b. The mean S/N per spectral bin is 11.  The data are normalized to 2.1265~$\mu$m and the error bars are derived from our analysis of injecting and extracting artificial planets (see $\S$\ref{sec:datared} and Figure~\ref{fig:fpsummary}).  Weak CH$_4$ absorption is suggested by the data from the negative slope between 2.15-2.22~$\mu$m, but strong 2.2~$\mu$m absorption is absent.  We note, however,  that the apparent dip at 2.18~$\mu$m is offset from the nominal 2.20~$\mu$m CH$_4$ bandhead.  \label{fig:finalspecfig} } 
\end{figure}

\begin{deluxetable}{lcc}
\tablewidth{0pt}
\tablecolumns{3}
\tablecaption{HR 8799 b Spectrum (OSIRIS/Kn3) \label{tab:spec}}
\tablehead{
        \colhead{$\lambda$ ($\mu$m)}   &    \colhead{$f_{\lambda}/f_{\lambda}$(2.1265 $\mu$m)}    &  \colhead{$\sigma_f$}
        }   
\startdata
2.1265     &     1.000     &    0.072    \\
2.1373     &     1.103     &    0.082    \\
2.1481     &     1.164     &    0.096    \\    
2.1589     &     1.004     &    0.058    \\    
2.1697     &     1.065     &    0.095    \\    
2.1805     &     0.820     &    0.067    \\
2.1913     &     0.778     &    0.098    \\    
2.2021     &     0.722     &    0.090    \\    
2.2129     &     0.697     &    0.085    \\    
2.2237     &     0.932     &    0.121    \\
\enddata
\end{deluxetable}

To telluric correct the spectrum of HR~8799~b we first extracted the spectrum of the standard star by spectrally binning the standard star data cubes, performing aperture photometry with an aperture radius of 10 spaxels (the FWHM is $\sim$2.4 spaxels), and combining the individual standard spectra in the same fashion as we did for the planet.  We then used the \texttt{xtellcor\_basic} package in Spextool (\citealt{Cushing:2004p501}) to divide the planet spectrum by the standard spectrum and multiply by a blackbody with a temperature of 10,000 K (corresponding to an A0 star) to restore the continuum shape.

Our telluric-corrected spectrum is displayed in Figure \ref{fig:finalspecfig} and the measurements are listed in Table \ref{tab:spec}.    Methane can influence the 2.15-2.23~$\mu$m spectral region, with a particularly prominent absorption feature at 2.20~$\mu$m (\citealt{Cushing:2005p288}).  Strong 2.2~$\mu$m CH$_4$ absorption is noticeably absent in our spectrum, although a weak methane feature may be present as a negative slope from 2.15-2.22~$\mu$m and diminished flux from 2.20-2.22~$\mu$m.

\subsection{Determining Measurement Errors}\label{sec:erranal}

\begin{figure}
  \resizebox{3.5in}{!}{\includegraphics{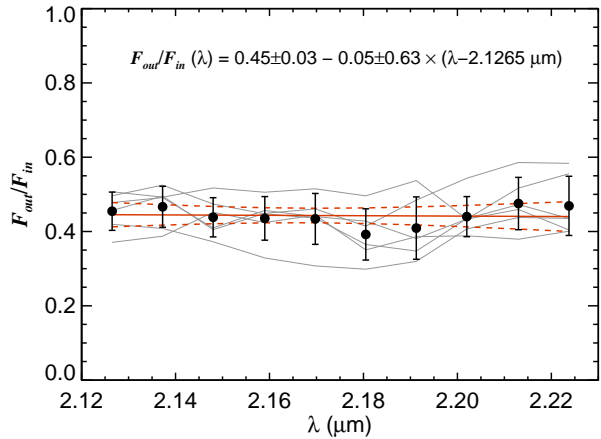}}
  \caption{Results of injecting and extracting six artificial planets in our data.  The fake planets were constructed from the standard star observations, scaled to the amplitude of HR~8799~b, inserted into all the individual images, and reduced using the LOCI algorithm.  The ratio of the input to output flux are shown at each spectral bin (gray lines).  A weighted linear fit to the mean and rms of the spectra (filled circles with error bars) yield a best-fitting slope consistent with zero (red solid line), suggesting that no systematic correction to the extracted spectrum of HR~8799~b is necessary.  The 1~$\sigma$ confidence interval for the linear fit is shown as red dashed curves, and the equation displays the best-fitting slope, offset, and parameter uncertainties.  A reduction in the extracted flux ($F_{out}/F_{in}$) is a characteristic of the LOCI algorithm (see \citealt{Lafreniere:2007p17998}).   \label{fig:fpsummary} } 
\end{figure}

Many sources of random and systematic errors can contribute to the uncertainties in our OSIRIS spectrum, including speckle noise, the use of a fixed radius in aperture photometry, and self-subtraction of the science target as a result of ADI+LOCI (e.g., \citealt{Lavigne:2009p19386}).  To assess the systematic and random errors in our spectrum, we injected and extracted fake planets in our data.  For the artificial planets we used the standard star scaled to the amplitude of HR 8799 b.  We injected the fake planets in six locations in the original individual images, reduced the full data set, and extracted the planets in the same fashion as we did with the science spectrum.  To minimize the impact on the construction of the reference PSF, we input one fake planet into the data at a time. The ratio of the input to output spectra of the fake planets varied on the level of $\sim$5-10\% depending on the location of the injected planet (Figure~\ref{fig:fpsummary}).  In some locations systematic trends as a function of wavelength were observed, especially in the region closer to HR 8799 where the speckle subtraction was worse.  Overall, however, a linear fit to the mean and rms of all six ratio spectra indicate that no correction is warranted as the slope of the fit is statistically consistent with zero.

When random noise is the dominant source of error, the best estimate of the actual flux at each wavelength is the mean value, and the best estimate of the uncertainty in that measurement is the standard error of the mean ($\sigma_M=\sigma$/$\sqrt N$, where $N$ is the number of spectra).  Here we use the median flux value at each wavelength instead of the mean to avoid outliers in the data.  In this case the expected uncertainty from photon noise is the standard error of the median (which for large sample sizes and normally distributed data is $\sqrt(\pi/2) \times \sigma$/$\sqrt N$ $\sim$ 1.25~$\sigma_M$; \citealt{Hojo:1931p315}).  However, our analysis of the fake planet spectra indicate that the  standard error of the median of the real planet spectra underestimates  the actual measurement uncertainty.

  To compute the uncertainties from our fake planet simulations, we first
  computed the spectra for each of the six fake planets by taking the mean
  spectrum of 18 measurements from the individual images. After the
  spectra of the 6 fake planets were scaled to a common level, the
  standard deviation of the six spectra at every wavelength was adopted as
  the true standard error of the mean. When we compared these
  uncertainties from the fake planets to the standard error of the median
  from the 18 real spectra of HR 8799 b (and accounting for the factor of
  1.25 between the two datasets), we found that the uncertainty derived
  from the spectra of HR 8799 b was underestimated by a factor of 1.5--2.5.
  We therefore adopted the standard error of the mean from the fake planet
  analysis for our HR 8799 b spectroscopic uncertainties. The final S/N
  per spectral bin of our OSIRIS spectrum varies between 7-18 with a mean
  value of 11.

\section{Comparison to Field Brown Dwarfs}\label{sec:bdcomp}

In many ways brown dwarfs are massive analogs of giant planets, sharing similar radii, effective temperatures, atmospheric physics, and cooling histories.  Brown dwarfs can therefore serve as natural reference objects to guide our understanding of giant planet properties.  Here we compare both our OSIRIS spectrum and previously published photometry of HR 8799 b to near-infrared spectra of L and T dwarfs from the Infrared Telescope Facility (IRTF) SpeX Prism Spectral Library.\footnote{Maintained by Adam Burgasser at http://www.browndwarfs.org/spexprism.}  The spectra have resolving powers between 75 and 120 and were reduced and telluric corrected using Spextool data reduction package for IRTF (\citealt{Vacca:2003p497}; \citealt{Cushing:2004p501}).  When available we use optical spectral types for L dwarfs, and for all other objects we use near-infrared types.  Objects with peculiar near-infrared classifications are omitted unless they have normal L-type optical spectral types.  Altogether we use 238 L and T dwarf spectra in our analysis, consisting of 92 objects with L-type optical classifications, 56 with L-type near-infrared classifications, and 90 with T-type near-infrared classifications.

\begin{figure}
  \vskip -.5 in
    \hskip -.7 in
  \resizebox{5in}{!}{\includegraphics{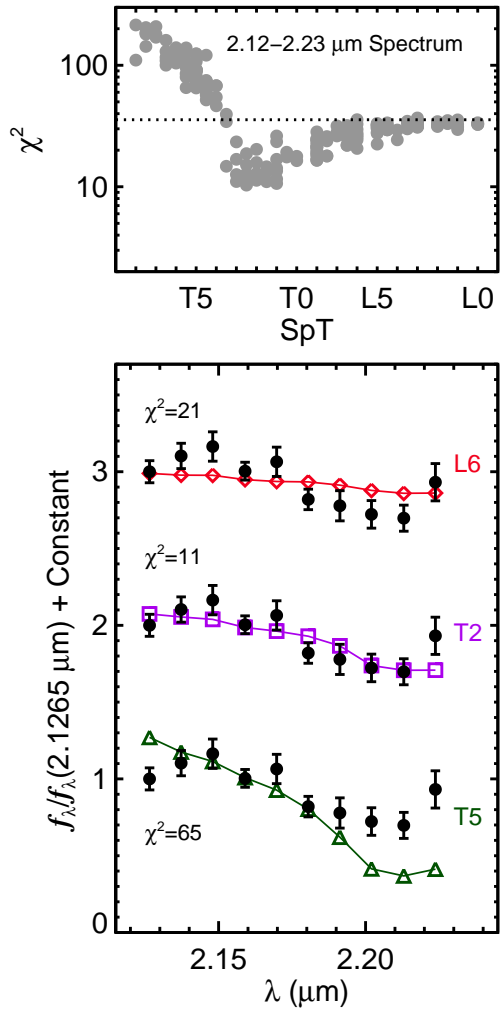}}
  \vskip -.5 in
  \caption{Results of fitting 238 field L and T dwarf spectra to our OSIRIS spectrum of HR~8799~b.   \emph{Top}: $\chi^2$ values as a function of spectral type.  The best-fitting spectral type is T2, although objects with spectral types earlier than T4 are consistent with the spectrum of HR~8799~b (see $\S$\ref{sec:bdcomp}).  Objects below the dotted line ($\chi^2$=35.6) are consistent with our data at the 3~$\sigma$ level.  \emph{Bottom}: Best-fitting spectra of field L6, T2, and T5 dwarfs compared to our OSIRIS spectrum.  From top to bottom the spectra represent 2MASS~J1036530--344138 (\citealt{Burgasser:2010p20068}), SDSS~J075840.33+324723.4 (\citealt{Burgasser:2008p14471}), and 2MASS~J0755480+221218  (\citealt{Burgasser:2006p14572}) with reduced $\chi^2$ values (9 degrees of freedom) of 2.30, 1.26, and 7.23, respectively.      \label{fig:bdcompfig_1} } 
\end{figure}

\subsection{2.12-2.23~$\mu$m Spectroscopy}

We fit the L and T dwarf prism spectra to our OSIRIS spectrum by  binning them onto the same wavelength grid as the OSIRIS data and scaling each prism spectrum $k$ by a factor $C_k$, which is found by minimizing the $\chi^2$ statistic

\begin{equation}\label{eq:chi2}
\chi^2_k = \sum_{i=1}^{n}{ \frac{(f_i - C_k \mathcal{F}_{k,i})^2} { \sigma_{f,i}^2} },
\end{equation}

\noindent where $f_i$ and $\sigma_{f,i}$ are the measured flux density and uncertainty of HR~8799~b and $\mathcal{F}_{k,i}$ is the prism spectrum flux density measurement at each wavelength $i$.  Equating the derivative to zero and solving for the scaling factor we get

\begin{equation}\label{eq:ck}
C_k = \frac{ \sum f_i \mathcal{F}_{k,i}/\sigma_{f,i}^2}{\sum \mathcal{F}_{k,i}^2/\sigma_{f,i}^2}.
\end{equation}

The best-fitting spectral type to our HR~8799~b spectrum is T2 (Figure \ref{fig:bdcompfig_1}).  Objects between L8 and T3 yield relatively small $\chi^2$ values ($<$20 for 9 d.o.f.).  Those later than T4 produce poor fits because of their strong methane absorption.  We use the $\Delta \chi^2$ method (\citealt{Press:2007p13558}) to determine which templates are statistically consistent with our data.   We consider  models with $\chi^2$ values $<$35.6 as being in agreement with our spectrum, which corresponds to a $\Delta$$\chi^2$ value with an integrated probability of 99.73\% (3~$\sigma$) for 9 degrees of freedom (25.3) plus the minimum $\chi^2$ value of the fits (10.3).  This corresponds to  spectral types earlier than T4.  For an integrated probability of 68.3\% ($\Delta$$\chi^2$ for 9 d.o.f. = 10.4), our OSIRIS data are consistent with spectral types of $\sim$L4-T3.\footnote{The best-fitting object results in a reduced $\chi^2$ value ($\chi^2$/$\nu$, for $\nu$=9 degrees of freedom) of 1.15 using the standard error of the mean uncertainties from the fake planet analysis.  The reduced $\chi^2$ value using the standard error of the median uncertainties from the 18 scaled spectra of HR~8799~b result in a reduced $\chi^2$ value of 2.39, which is further evidence that adopting the standard errors from the extracted spectra underestimate the uncertainties (see $\S$\ref{sec:erranal}).}

\begin{figure*}
  \vskip -.5 in
    \hskip 1.1 in
  \resizebox{4.5in}{!}{\includegraphics{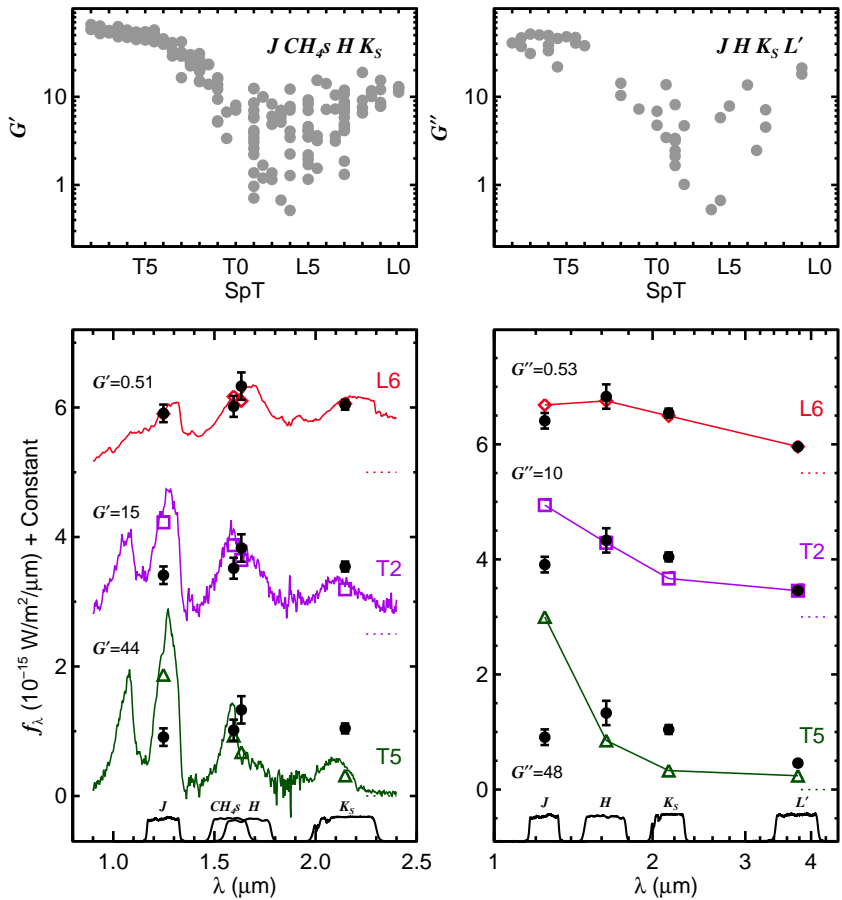}}
  \vskip -.6 in
  \caption{Empirical comparison of field L and T dwarfs to 1-4~$\mu$m photometry of HR~8799~b from \citet{Marois:2008p18841}.  \emph{Top Left}: Goodness-of-fit statistic $G'$ (Equation \ref{eq:gstat}) as a function of spectral type for fits to the $J$, $CH_4s$, $H$, and $K_S$ bands.  Photometry is synthesized from 238 objects in the SpeX Prism Spectral Library.  Ultracool dwarfs with spectral types of L5-L8 provide the best fits.  See Figure~\ref{fig:bdcompfig_3} for the best-fitting individual objects.  \emph{Bottom Left}: Best-fitting L6, T2, and T5 spectra of field dwarfs scaled to the photometry of HR~8799~b at 10 pc.  Mid- to late-T dwarfs provide poor fits to the NIR SED of HR~8799~b as a result of  brighter $J$-band and fainter $K$-band flux.  The spectra represent 2MASS~J21481628+4003593 (L6, red; \citealt{Looper:2008p14686}), SDSS~J143553.25+112948.6 (T2, blue; \citealt{Chiu:2006p17707}), and 2MASS~J23312378--4718274 (T5, green; \citealt{Burgasser:2004p574}). \emph{Top Right}: Goodness-of-fit statistic $G''$ (Equation~\ref{eq:gstat2}) as a function of near-infrared spectral type for fits of 40 ultracool dwarfs with $J$-, $H$-, $K_S$-, and $L'$-band photometry from the compilation in \citet[their Table 3]{Leggett:2010p20094}. Mid- to late-L dwarfs fit the 1-4~$\mu$m photometry of HR~8799~b the best.  \emph{Bottom Right}: Best-fitting L6, T2, and T5 objects.  From top to bottom the spectra represent 2MASS~J08251968+2115521, SDSS~J075840.33+324723.4, 2MASS~J15031961+2525196 (for details see references in Table 3 of \citealt{Leggett:2010p20094}).\label{fig:bdcompfig_2} } 
\end{figure*}

\subsection{1.1-2.4~$\mu$m Photometry}

We perform a similar analysis using previously published photometry of HR~8799~b.  \citet{Marois:2008p18841} presented $J$, $CH_{4}s$, $H$,  $CH_{4}l$, $K_S$, and $L'$ photometry of HR~8799~b in their discovery paper.  \citet{Lafreniere:2009p17982} published $HST$ $F160W$ photometry, \citet{Fukagawa:2009p18543} and \citet{Metchev:2009p19676} $H$-band photometry, and \citet{Hinz:2010p20424} an $L'$ detection with upper limits at 3.3~$\mu$m and $M$-band.  We use only the Marois et al. $J$, $CH_{4}s$, $H$, and $K_S$ photometry in this analysis to avoid overweighting the $H$-band region of the spectrum in our fits.

We synthesize photometry (e.g., Equation 5 of \citealt{Tokunaga:2005p18542}) from each prism spectrum of the field dwarf sample using the $J$, $H$, and $K_S$ bands from the Mauna Kea Observatory (MKO) filter consortium (\citealt{Simons:2002p20490}; \citealt{Tokunaga:2002p20495}) and the zero point Vega flux densities from \citet{Tokunaga:2005p18542}.  For $CH_{4}s$, we use the Keck~II/NIRC2 filter transmission curve (J.~Lyke, private communication) and the flux-calibrated Vega model spectrum from Spextool (\citealt{Cushing:2004p501}) to compute a zero point flux density for that filter (5.31$\times$10$^{-11}$~W/m$^2$/$\mu$m).

We use a modified form of the $\chi^2$ statistic to assess goodness-of-fits for our photometric comparison.  Following \citet{Cushing:2008p2613} we compute a $G^{'}$ statistic for each prism spectrum $k$:

\begin{equation}\label{eq:gstat}
G_k^{'} = \sum_{i=1}^{n}w_i{ \frac{(f_i -  C_k^{'} \langle \mathcal{F}_{k,i} \rangle )^2} { \sigma_{f,i}^2  }},
\end{equation}

\noindent where $w_i$ is the weight applied to each photometric point $i$ and  $\langle \mathcal{F}_{k,i} \rangle$ is the monochromatic flux density in each bandpass.  The scaling factor is then

\begin{equation}\label{eq:gstatck}
C_k^{'} = \frac{ \sum w_i f_i \langle \mathcal{F}_{k,i} \rangle/\sigma_{f,i}^2}{\sum w_i \langle \mathcal{F}_{k,i} \rangle^2/\sigma_{f,i}^2}.
\end{equation}

\noindent We experimented with equal weights ($w_i$=1.0) for each filter, in which case $G_k^{'}$ becomes the usual $\chi^2_k$ statistic, and weights proportional to the width of each spectral bandpass such that $\sum w_i$=1, defined as 

\begin{equation}\label{eqn:weights}
w_i = \frac{\Delta \lambda_i}{ \sum_{j=1}^{n} \Delta \lambda_j},
\end{equation}

\noindent  where $n$ is the number of filters used in the fit.  This  results in weights of 0.18, 0.33, 0.13, and 0.36 for the $J$, $H$, $CH_{4}s$, and $K_S$ bands.   We use this weighting scheme in our analysis, although using equal weights produces similar results.

The results of our photometric comparison are shown in the left panels of Figure \ref{fig:bdcompfig_2}.  Overall, field brown dwarfs produce poor fits to the HR~8799~b photometry.  This may be a result of comparing objects with an order of magnitude difference in surface gravity (log~$g$$\sim$4.0--4.5 [cgs] for HR~8799~b vs. log~$g$$\sim$5.0--5.5 for field brown dwarfs).  The differences can also arise from comparing objects with different cloud properties and/or metallicities.  T dwarfs produce the worst fits as a result of their blue $J$--$K$ colors compared to HR~8799~b.  The best-fitting \emph{spectral type} (as measured by the average $G'$-value for a given spectral type) is L5, although similarly good $G'$ values are obtained from L5 to L8.  However, typical L5 objects with $G'$ values near the mean for that spectral type do not produce a strong match because of the red colors of HR~8799~b compared to other field objects (see also Figure 6 of \citealt{Allers:2010p20499}).  

\begin{figure*}
  \begin{center}
  \resizebox{4.5in}{!}{\includegraphics{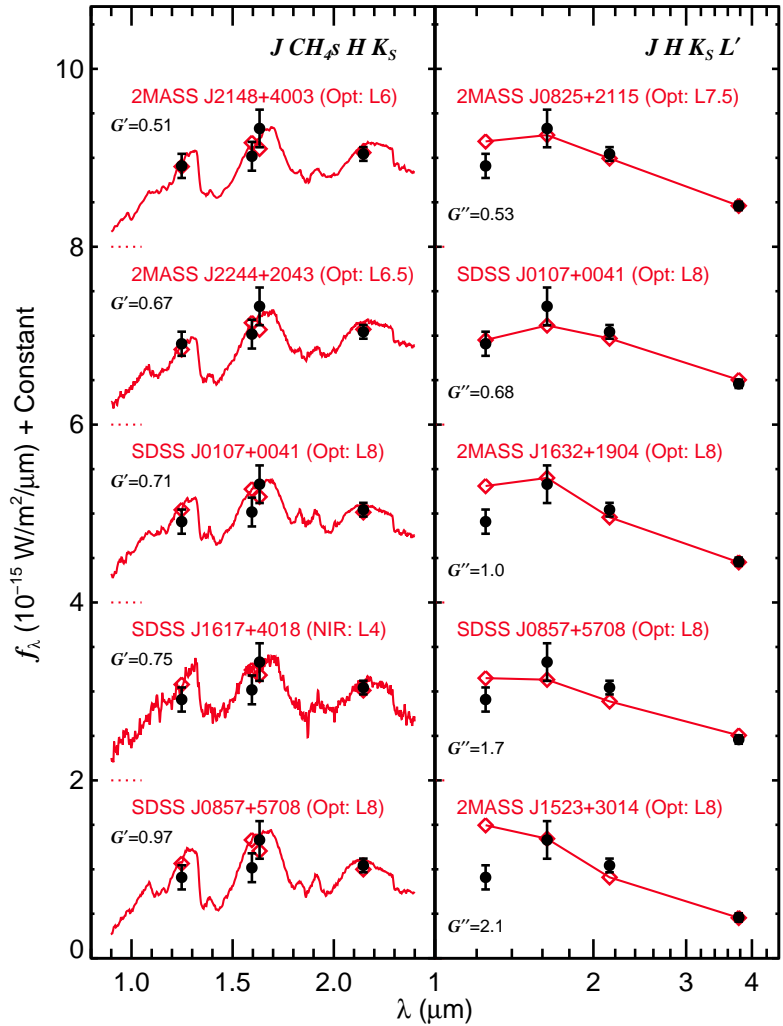}}
  \end{center}
  \vskip -.4 in
  \caption{\emph{Left}: Best-fitting objects to $J$-, $CH_4s$-, $H$-, and $K_S$-band photometry of HR~8799~b (at 10 pc; see also Figure~\ref{fig:bdcompfig_2}).  These ultracool dwarfs generally have late-L optical spectral types, red colors compared to normal field dwarfs with the same spectral types, and indications of low surface gravitites and/or cloudy atmospheres.  Although the NIR SEDs provide good matches, the luminosities of these field objects are $\sim$3-10 times higher than that of HR~8799~b.  The SpeX/prism spectra of 2MASS~J21481628+4003593 and 2MASS~J2244316+204343 were originally published in \citet{Looper:2008p14686}, SDSS~J010752.33+004156.1 and SDSS~J0857+5708 in \citet{Burgasser:2010p20068}, and SDSS~J161731.65+401859.7 in \citet{Chiu:2006p17707}.  \emph{Right}: Best-fitting objects to the 1-4~$\mu$m photometry of HR~8799~b.  See $\S$\ref{sec:bdcomp} and Figure~\ref{fig:bdcompfig_2} for details about the fitting procedure.\label{fig:bdcompfig_3} } 
\end{figure*}

The best fitting \emph{individual} objects produce much better matches to the HR~8799~b NIR photometry.  The left panel of Figure \ref{fig:bdcompfig_3} shows the five best-fitting ultracool dwarfs, which have optical spectral types between L6 and L8 (for those with optical classifications).  These objects all have very red NIR colors ($J$--$K_S$$\sim$2.0--2.5~mag) and generally show evidence of low gravity and/or thick clouds.   For example, atmospheric model fitting to the 0.8-14.5~$\mu$m spectrum of SDSS~J085758.45+570851.4 indicates that it has a low surface gravity (log~$g$=4.5) and a very cloudy atmosphere (\citealt{Stephens:2009p19484}).  HR~8799~b is well fit by 2MASS~J22443167+2043433 (2MASS~J2244+2043), which is the reddest L~dwarf known with a $J$--$K_S$ color of 2.48$\pm$0.15~mag (\citealt{Dahn:2002p13692}).  A comparison of atmospheric models to mid-infrared photometry of 2MASS~J2244+2043 presented by \citet{Leggett:2007p18996} suggests that it has extreme cloud properties with strong vertical mixing ($K_{zz}$$>$10$^4$~cm$^2$~s$^{-1}$).  There is also evidence that 2MASS~J2244+2043 has a low surface gravity based on weak \ion{K}{1} lines and FeH bands (\citealt{McLean:2003p3912}; \citealt{Looper:2008p14686}).  The overall best-fitting object to HR~8799~b is 2MASS~J21481628+4003593, which has a $J$--$K_S$ color of 2.38$\pm$0.06~mag and a triangular $H$-band shape (\citealt{Looper:2008p14686}; \citealt{Allers:2007p66}).  However, \citet{Looper:2008p14686} interpret this object as old and metal-rich.  Note that although the NIR SED of HR~8799~b is well-matched by these red field L dwarfs, the luminosity of HR~8799~b is roughly 3-10 times lower than these best-fitting objects.

\subsection{1.1-4.1~$\mu$m Photometry}

We now extend the empirical comparison to $4$~$\mu$m by incorporating $L'$-band photometry.  We use the compilation of ultracool dwarf near- and mid-infrared photometry from \citet{Leggett:2010p20094} to fit the $J$, $H$, $K_S$, and $L'$ photometry of HR~8799~b.  $K_\mathrm{MKO}$ magnitudes are converted to $K_S$ magnitudes using a polynomial fit to differenced synthetic photometry of L and T dwarfs from the SpeX Spectral Prism Library (see Appendix~\ref{app:kmkotoks}).  Uncertainties in the polynomial fit are accounted for in the $K_S$-band photometric error.

To incorporate the photometric measurement errors of the comparison objects into our fitting analysis we define a goodness-of-fit statistic $G''$ as follows:

\begin{equation}\label{eq:gstat2}
G_k^{''} = \sum_{i=1}^{n}w_i{ \frac{(f_i -  C_k^{''}  \mathcal{F}_{k,i}  )^2} { \sigma_{f,i}^2 + (C_k^{''} \sigma_{ \mathcal{F}_{k,i} })^2  } },
\end{equation}

\noindent where $\sigma_{  \mathcal{F}_{k,i} }$ is the photometric measurement uncertainty of the ultracool dwarf $k$ for filter $i$.\footnote{We remove angle brackets from $\mathcal{F}_{k,i}$ in Equation~\ref{eq:gstat2} to distinguish between actual photometry and synthesized photometry from spectra, which we use in Equation~\ref{eq:gstat}.}  The scaling factor $C_k^{''}$ is calculated iteratively as described in Appendix~\ref{app:gstatckdp}.  The weights $w_i$ are defined as in Equation~\ref{eqn:weights} and correspond to values of 0.11, 0.19, 0.22, and 0.48 for the $J$, $H$, $K_S$, and $L'$ filters.

Mid- to late-L dwarfs provide the best fits to the 1-4~$\mu$m photometry of HR~8799~b (Figure~\ref{fig:bdcompfig_2}, right panels).  These results hold when the weights are set to unity for all bands.  Similar to our empirical comparison of the NIR SED (1.1-2.4~$\mu$m), the best-fitting field objects to the near- and mid-IR photometry (1.1-4.1~$\mu$m) are late-L dwarfs (L7.5-L8 optical spectral types) with red NIR colors, most of which exhibit evidence of low gravities and/or abnormally dusty photospheres (Figure~\ref{fig:bdcompfig_3}, right panel).  2MASS~J08251968+2115521 (hereafter 2MASS~J0825+2115) has the lowest $G''$ value and is a well-studied red L dwarf ($J$--$K_S$=2.07$\pm$0.04 mag; L7.5 optical spectral type).  \citet{Stephens:2009p19484} and \citet{Cushing:2008p2613} performed atmospheric model fits to its near- and mid-infrared spectrum and found best-fitting effective temperatures of 1200~K and 1400~K, respectively, with unusually thick clouds compared to other field L dwarfs.

\begin{figure}
  \begin{center}
  \resizebox{3.5in}{!}{\includegraphics{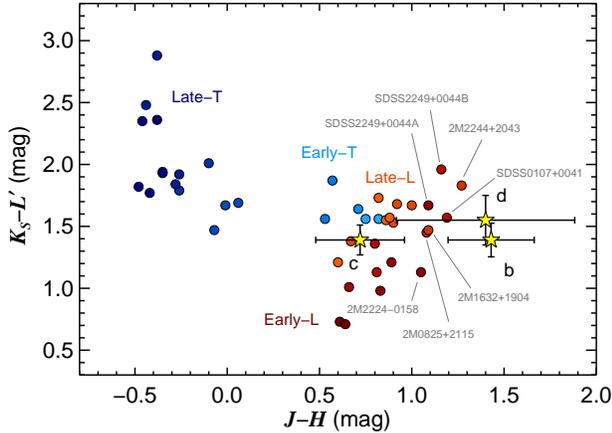}}
  \end{center}
  \caption{Color color diagram for field L and T dwarfs from \citet{Leggett:2010p20094} compared to the HR~8799 planets. L~dwarfs are plotted as red (early-L) to orange (late-L) and T dwarfs are plotted as light blue (early-T) to dark blue (late-T).  The HR~8799 planets are shown as yellow stars and objects with red $J$--$H$ colors are labelled. HR~8799~b is redder in $J$--$H$ than field L and T dwarfs (the large  uncertainty in $J$--$H$ for planet d is consistent with late-L dwarfs).  HR~8799~c shares similar colors to field objects near the L/T transition.  In addition to objects from  \citet{Leggett:2010p20094}, we also overplot the position of the recently discovered young L~dwarf binary SDSS~J224953.47+004404.6AB (\citealt{Allers:2010p20499}; $K$-band magnitudes are converted to $K_S$ using the relation in Appendix~\ref{app:kmkotoks}). \label{fig:jhklcc} } 
\end{figure}

The location of HR~8799~b in the ($J$--$H$, $K_S$-$L'$) color-color diagram is shown in Figure~\ref{fig:jhklcc}.  HR~8799~b has a very red $J$--$H$ color (1.43 mag) compared to field L and T dwarfs.  Its $J$--$H$ color is even more extreme than the reddest field L dwarfs (labelled in Figure \ref{fig:jhklcc}), which are generally thought to have low gravities and/or unusually dusty photospheres.

\begin{figure}
  \begin{center}
  \resizebox{3.5in}{!}{\includegraphics{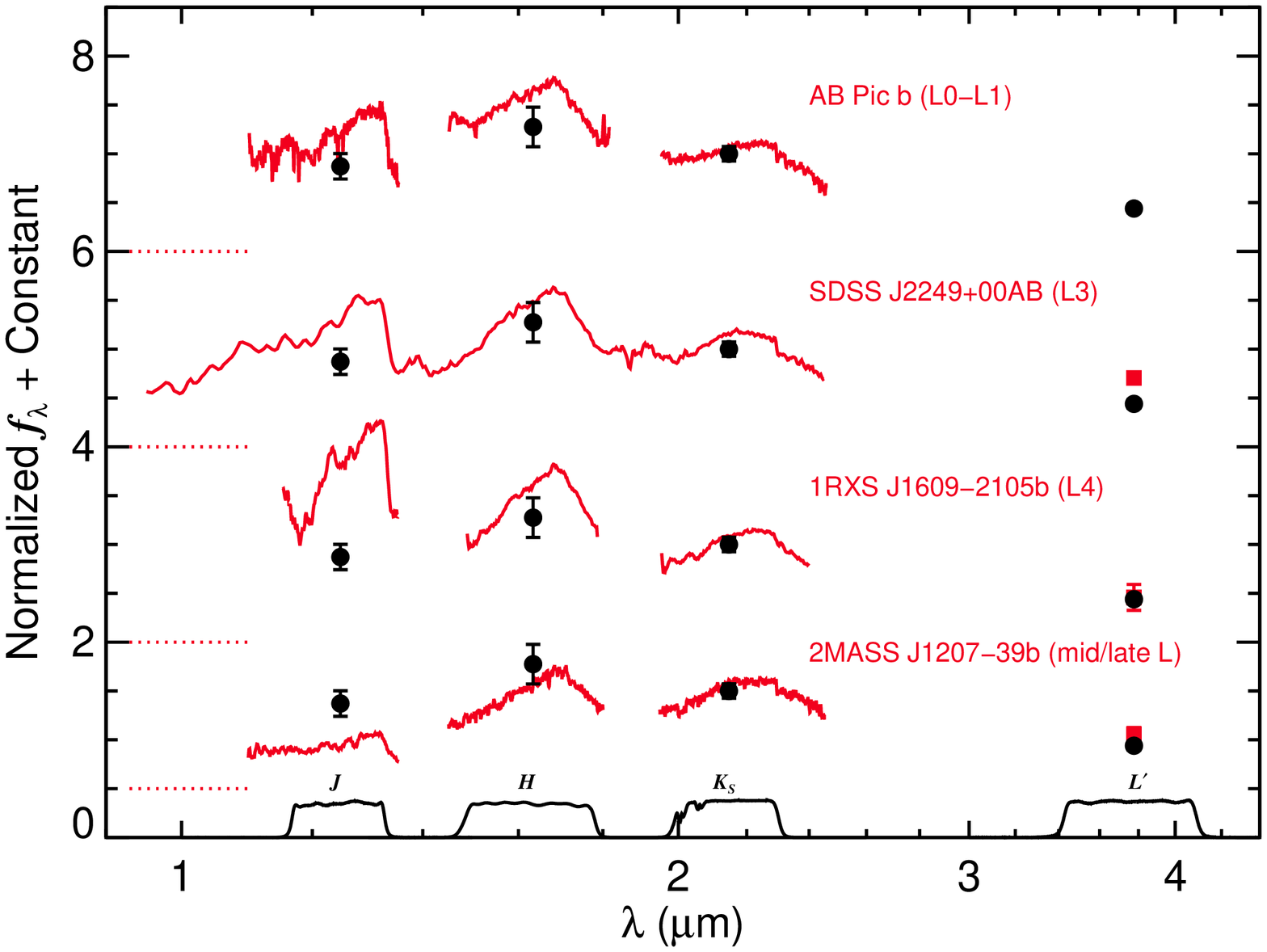}}
  \end{center}
  \caption{1.1-4.1~$\mu$m photometry of HR~8799~b (black circles) compared to four well-studied, low-mass, low-gravity L dwarfs (red).  The spectra are normalized between 2.0-2.4~$\mu$m and are offset by a constant, and the HR~8799~b SED is normalized to the $K_S$ band.  The NIR ($\sim$1.1-2.5~$\mu$m) spectra of AB~Pic~b (\citealt{Bonnefoy:2010p20602}), SDSS~J2249+0044AB (\citealt{Allers:2010p20499}), and 1RXS~J1609--2105 (\citealt{Lafreniere:2008p14057}; \citealt{Lafreniere:2010p20763}) are bluer than HR~8799~b, although the NIR SED of 2MASS~J1207-3932~b is redder.  $L'$-band photometry is included for SDSS~J2249+0044AB (\citealt{Allers:2010p20499}), 1RXS~J1609--2105b (\citealt{Lafreniere:2010p20763}), and 2MASS~J1207-3932~b (\citealt{Mohanty:2007p6975}).    None of these low-gravity L dwarfs provide good matches to the SED of HR~8799~b.  \label{fig:famous_young_ldwarf_comp} } 
\end{figure}

In Figure~\ref{fig:famous_young_ldwarf_comp} we compare the 1.1-4.1~$\mu$m photometry of HR~8799~b to four well-studied, young, low-mass L dwarfs.  AB~Pic~b is a $\sim$13-14~$M_\mathrm{Jup}$ companion to a young ($\sim$30~Myr) K2-type member of the  Tucana-Horologium association (\citealt{Chauvin:2005p19642}).  Recently, \citet{Bonnefoy:2010p20602} obtained medium-resolution integral-field spectroscopy of AB~Pic~b from 1.1-2.5~$\mu$m and determined a spectral type of L0-L1.  A comparison of HR~8799~b to the Bonnefoy~et~al. NIR spectrum in Figure~\ref{fig:famous_young_ldwarf_comp} shows a decent match from 1.1-2.4~$\mu$m, but HR~8799~b is slightly redder in $J$--$K_S$.  Similarly, HR~8799~b has redder NIR colors compared to the integrated-light spectrum of SDSS~J224953.47+004404.6AB (hereafter SDSS~J2249+0044AB), a pair of low-gravity L~dwarfs (L3+L5) recently discovered by \citet{Allers:2010p20499}.  The NIR spectrum of the young $\sim$8~$M_\mathrm{Jup}$ L4$^{+1}_{-2}$-type companion 1RXS~J160929.1--210524~b from \citet{Lafreniere:2008p14057} and \citet{Lafreniere:2010p20763} is significantly bluer than HR~8799~b, although the $K_S$--$L'$ color is nearly identical.  Finally, HR~8799~b appears to be slightly bluer compared to the NIR spectrum of the $\sim$5-8~$M_\mathrm{Jup}$ mid/late L-type object 2MASS~J1207-3932~b from \citet{Patience:2010p20422}.  None of these four low-gravity L dwarfs provide good templates to the 1.1-4.1~$\mu$m photometry of HR~8799~b.

Altogether, our spectral and photometric comparisons to field brown dwarfs suggest a spectral type between L5 and T2 for HR~8799~b.  Although peculiar compared to most L and T dwarfs in the field, the planet's photometry is consistent with the reddest field L dwarfs.  These results imply that HR~8799~b is the lowest-mass L/T transition object currently known.

\section{Comparison to Atmospheric Models}\label{sec:atmmod}

\begin{figure*}
  \begin{center}
  \resizebox{5.5in}{!}{\includegraphics{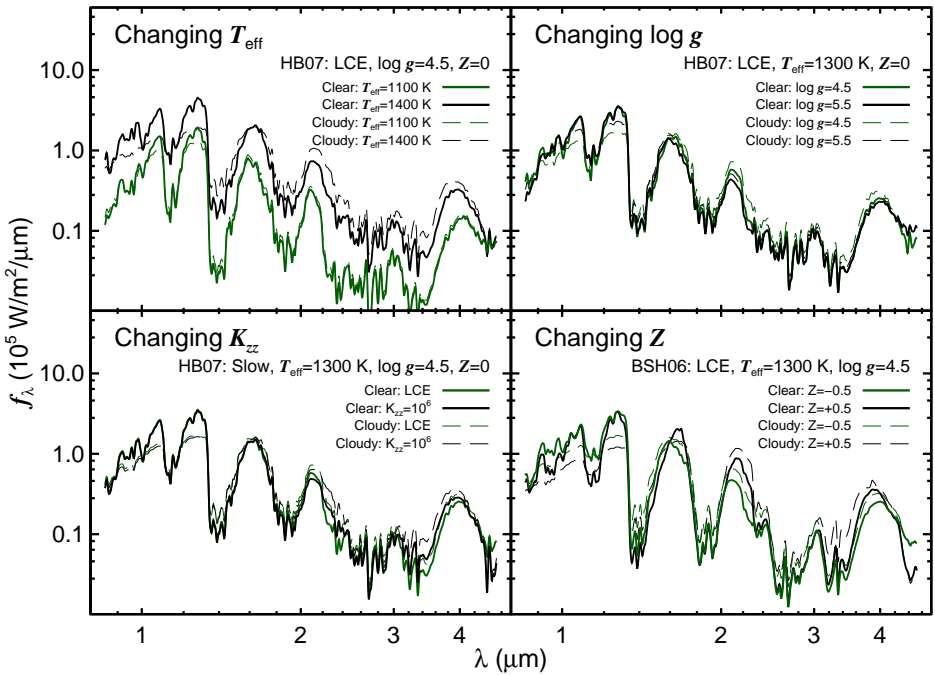}}
  \end{center}
  \caption{Atmospheric models from HB07 and BSH06 for changing $T_\mathrm{eff}$, log~$g$, eddy diffusion coefficient ($K_{zz}$), and metallicity ($Z$) for clear and cloudy atmospheres.  Clouds  (thin dashed lines) produce dramatic effects on emergent spectra  by suppressing the $J$-band flux and enhancing the $K$-band flux.  The shape and amplitude of the 2.2~$\mu$m region can be influenced by all four parameters.    \label{fig:burrows_plot_vary} } 
\end{figure*}

To derive the physical properties of HR~8799~b we compare our 2.12-2.23~$\mu$m spectrum and 1.1-4.1~$\mu$m published photometry to atmospheric models.  We use three published grids of low-temperature atmospheric models in our analysis.  The models of \citet[hereafter HB07]{Hubeny:2007p14693} were developed to assess the effects of nonequilibrium chemistry on the emergent spectra of brown dwarfs and giant planets.  In these objects, departure from local chemical equilibrium (LCE) can arise when vertical mixing dredges up molecules from warmer, deeper atmospheric layers on timescales shorter than chemical reaction timescales and primarily affects CO/CH$_4$ and N$_2$/NH$_3$ chemical abundances.  The HB07 grid includes a baseline LCE model as well as three variations of the mixing timescale, which is parametrized by the eddy diffusion coefficient $K_{zz}$ (with model values of 0, 10$^2$, 10$^4$, and 10$^6$ cm s$^{-1}$).  HB07 compute models for three CO/CH$_4$ chemical reaction timescales (``$slow$,'' ``\emph{fast1},'' and ``\emph{fast2}'').  We consider the $slow$ and \emph{fast2} chemical timescale prescriptions (\emph{fast1} and \emph{fast2} are similar) spanning effective temperatures of 700--1800~K ($\Delta T_\mathrm{eff}$=100~K; see \citealt{Hubeny:2007p14693} for details).  The models are for solar metallicity abundances and include clear atmospheres and cloudy atmospheres containing 100~$\mu$m forsterite particles (Figure \ref{fig:burrows_plot_vary}).  The surface gravity of HR~8799~b is expected to be $\sim$4.0-4.7~dex given its luminosity and an age range of 30-160 Myrs (\citealt{Baraffe:2003p588}), so we use the log~$g$=4.5 models here.

Metallicity plays an important role in shaping the near-infrared spectra of ultracool objects by impacting the strength of collision-induced absorption by H$_2$ (CIA H$_2$; \citealt{Linsky:1969p3947}; \citealt{Borysow:1997p3945}).  CIA H$_2$ is strongest in the 1.3--2.5~$\mu$m spectral region and can affect the shape and amplitude of the $K$-band emergent spectrum; metallicity variations are therefore directly relevant to our OSIRIS spectrum.  We use the grid of low-temperature LCE atmospheric models from \citet[hereafter BSH06]{Burrows:2006p7009}, which sample an order of magnitude variation in metallicity ($Z$=--0.5, 0.0, +0.5) for effective temperatures of 700-1800~K (Figure \ref{fig:burrows_plot_vary}).  We consider log~$g$=4.5 models with clear and cloudy variants for our analysis.  As emphasized in BSH06, we note that there remain many uncertainties in the meteorological physics behind cloudy models in general.  This particular set is only a representative version of particle size and cloud prescription which works well for modeling L dwarfs.

\begin{figure*}
    \begin{center}
  \resizebox{5.5in}{!}{\includegraphics{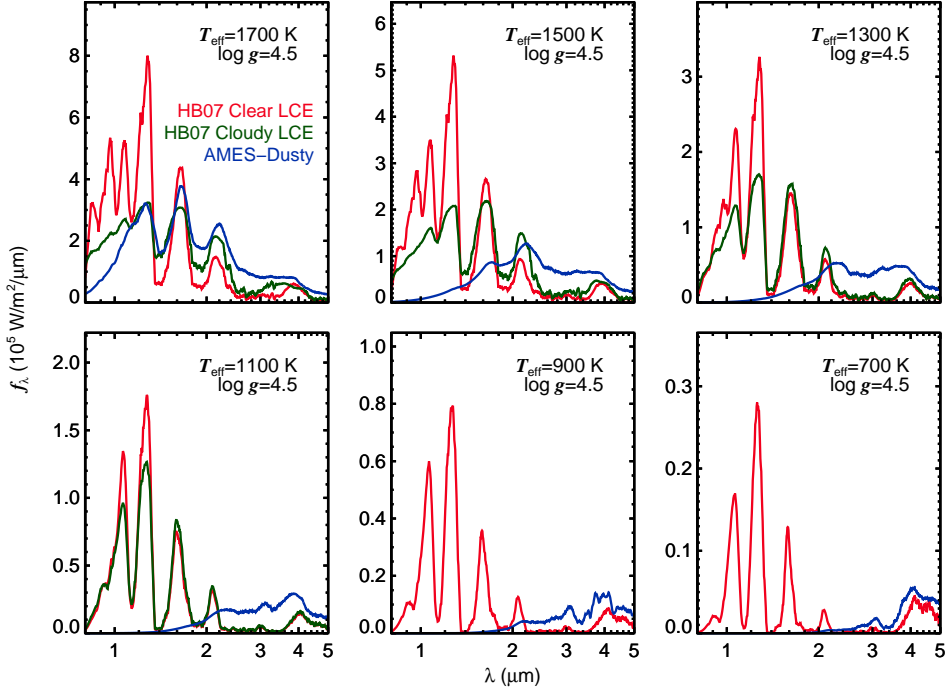}}
  \end{center}
  \caption{HB07 clear (red), HB07 cloudy (green), and Ames-Dusty (blue) synthetic spectra for log~$g$=4.5 and $T_\mathrm{eff}$=1700, 1500, 1300, 1100, 900, and 700~K.  Increasing the photospheric dust content suppresses flux at shorter wavelengths, which then reemerges at longer wavelengths.  At low effective temperatures the HB07 cloudy variant is similar to the clear version, but the extreme Ames-Dusty case is dramatically different with virtually no flux emitted in the $J$ and $H$ bands.  Note that the HB07 LCE cloudy models end at 1000~K and so are not plotted in the 900~K and 700~K panels.     \label{fig:bur_amesdusty_comp} } 
\end{figure*}

The red colors of HR~8799~b suggest that photospheric dust may play an important role in shaping its emergent spectrum.  The HB07 and BSH06 models we consider here include a single cloud prescription with modest dust content and large grain sizes.  To compare with a more extreme case of dust formation we also use the solar metallicity Ames-Dusty atmospheric models of \citet{Allard:2001p14776}, which considers the limiting case of dust formation with no gravitational settling and is computed in LCE.  Thirty types of spherical grains are included in the models with an interstellar size distribution from 0.00625-0.24~$\mu$m.  The grid contains effective temperatures from 500-2000~K ($\Delta$$T_\mathrm{eff}$=100~K) and surface gravities from 3.5-6.0 dex ($\Delta$log~$g$=0.5).  The effect of atmospheric dust is to increase the gas temperature in the outer photospheric layers compared to the dust-free case (\citealt{Allard:2001p14776}).  This heating can result in weaker molecular features (like CH$_4$ and H$_2$O) and, in combination with the depletion of metals from the gas and the higher grain opacities at shorter wavelengths, tends to smooth out the SED to more closely resemble that of a blackbody.  This phenomenon is shown in Figure~\ref{fig:bur_amesdusty_comp} in a comparison between the HB07 clear (red), HB07 cloudy (green), and Ames-Dusty models (blue), which progressively shows the impact of higher levels of photospheric dust at different effective temperatures. (Note that the LCE HB07 cloudy models end at 1000~K.)

\begin{figure}
\begin{center}
  \vskip -.285 in
  \resizebox{3.5in}{!}{\includegraphics{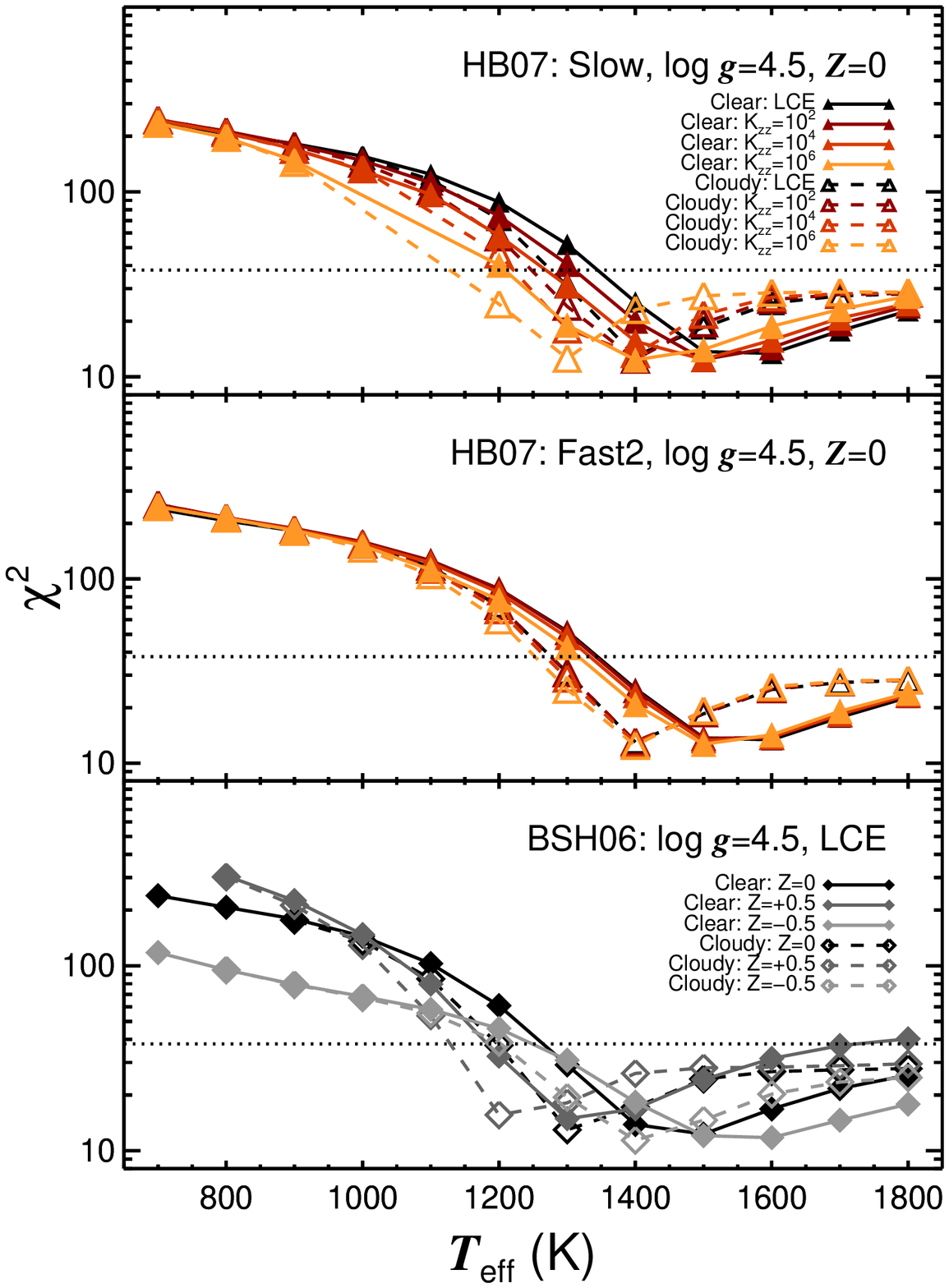}}
\end{center}
  \caption{Results from fitting atmospheric models to our $K$-band spectrum of HR~8799~b.  The $\chi^2$ values using the HB07 models are shown in the top and middle panels for the $slow$ and \emph{fast2} chemical reaction timescales.  The bottom panel displays the results using the BSH06 equilibrium models for three metallicities.  Models with clear atmospheres are shown with solid lines while those with cloudy atmospheres have dashed lines.  At low temperatures ($\lesssim$1200~K) the $\chi^2$ values increase as a result of stronger 2.2~$\mu$m CH$_4$ absorption in the models.  We consider models with $\chi^2$ values $<$37.8 (99.73\% confidence level) as being consistent with the data (dotted line; see $\S$\ref{subsec:specanal}).    \label{fig:chi2summaryfig} } 
\end{figure}

\subsection{Physical Properties from Spectroscopic Analysis}\label{subsec:specanal}

\begin{figure}
  \resizebox{3.5in}{!}{\includegraphics{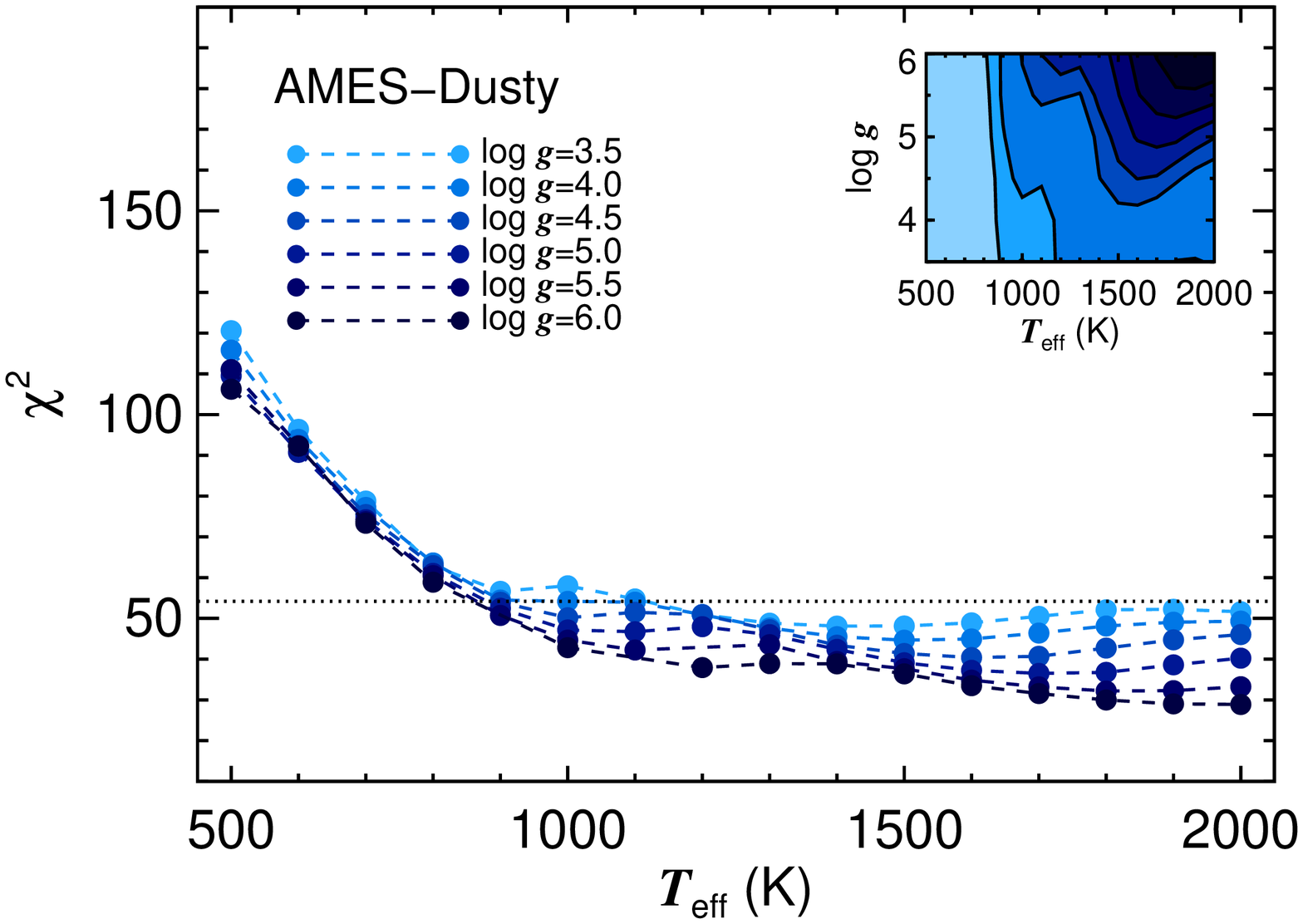}}
  \caption{Results of fitting the Ames-Dusty models to our $K$-band spectrum.  The greenhouse heating from the dust inhibits the formation of CH$_4$, so the 2.2~$\mu$m feature is absent in these models and the $\chi^2$ values remain similar over a wide range of temperatures ($\sim$1000-2000~K).  The inset displays the same data as a contour plot with contour levels of \{1.1,1.2,1.3,1.4,1.5,1.8,2.\} times the minimum $\chi^2$ value (28.9 at $T_\mathrm{eff}$=2000~K, log~$g$=6.0).  The dotted line shows the $\chi^2$ cutoff value for being consistent with the data at the 3~$\sigma$ confidence level.  The Ames-Dusty models imply an effective temperature $>$800~K for fits to our OSIRIS spectrum.  \label{fig:chi2_amesdusty} } 
\end{figure}

We use the same $\chi^2$ fitting procedure as we did for field brown dwarfs in $\S$\ref{sec:bdcomp} to compare the model atmosphere spectra to our OSIRIS spectrum.  Our OSIRIS spectrum is not flux calibrated so for our analysis of the spectrum we cannot compute a radius as we do in $\S$\ref{sec:photanal} using photometry.  For the same reason we do not include the spectrum into a joint analysis with the photometry.  The $\chi^2$ values for the HB07 and BSH06 models continue to decline with increasing $T_\mathrm{eff}$ until they reach a minimum at $\sim$1300-1600~K (Figure~\ref{fig:chi2summaryfig}).  The overall behavior of the $\chi^2$ values is a result of the strength of the 2.2~$\mu$m CH$_4$ feature in the models diminishing with higher temperature as the carbon balance shifts from CH$_4$ to higher CO abundances.  The result is a flattening the 2.12-2.23~$\mu$m spectral region in the models, producing better fits to our HR~8799~b spectrum.  In contrast, the $\chi^2$ distributions for the Ames-Dusty models (Figure~\ref{fig:chi2_amesdusty}) are relatively flat for effective temperatures higher than $\sim$800~K as a result of inhibited methane formation.

For effective temperatures between $\sim$800-1300~K, the HB07 $slow$ models with clouds and high $K_{zz}$ values produce better fits than clear LCE models.  The $\chi^2$ distributions for cloudy HB07 models result minima $\sim$100~K lower than for the clear models.  For the BSH06 models, the cloudy versions tend to fit the data slightly better than clear models.  Higher gravity Ames-Dusty models tend to fit the spectrum better than lower gravities for $T_\mathrm{eff}$$>$900~K, although the actual goodness-of-fit for the Ames-Dusty models is worse than for the HB07 and BSH06 models as is evident from the $\chi^2$ values.  

The dominant opacity sources (CH$_4$, CIA H$_2$, and H$_2$O) are generally well understood in the narrow spectral range of the observations, so systematic errors in the $K$-band region of the models are probably small compared to the entire spectral energy distribution. (See $\S$3.2 of \citealt{Bowler:2009p19621} and \citealt{Dupuy:2010p20924} for a discussion of this problem.)  We therefore use the $\Delta \chi^2$ method to assign confidence limits as we did for field brown dwarfs in $\S$\ref{sec:bdcomp}, except here we employ a $\chi^2$ value of 37.8 as a cutoff value between consistent and inconsistent models (corresponding to a 3~$\sigma$ confidence level) for the HB07 and BSH6 grids because the minimum $\chi^2$ value is 12.5.  For the Ames-Dusty grid (which we consider distinct enough to merit its own cutoff value), the equivalent $\chi^2$ value is 54.2 ($\chi^2_\mathrm{min}$=28.9). If HR~8799~b has parameters that fall within these grids, and if the atmospheric models are correct, the HB07 and BSH06 models constrain the effective temperature of HR~8799~b to $>$1100~K, and the Ames-Dusty models constrain the effective temperature to $>$800~K.

\begin{figure*}
\begin{center}
  \vskip -1.7 in
  \resizebox{7.5in}{!}{\includegraphics{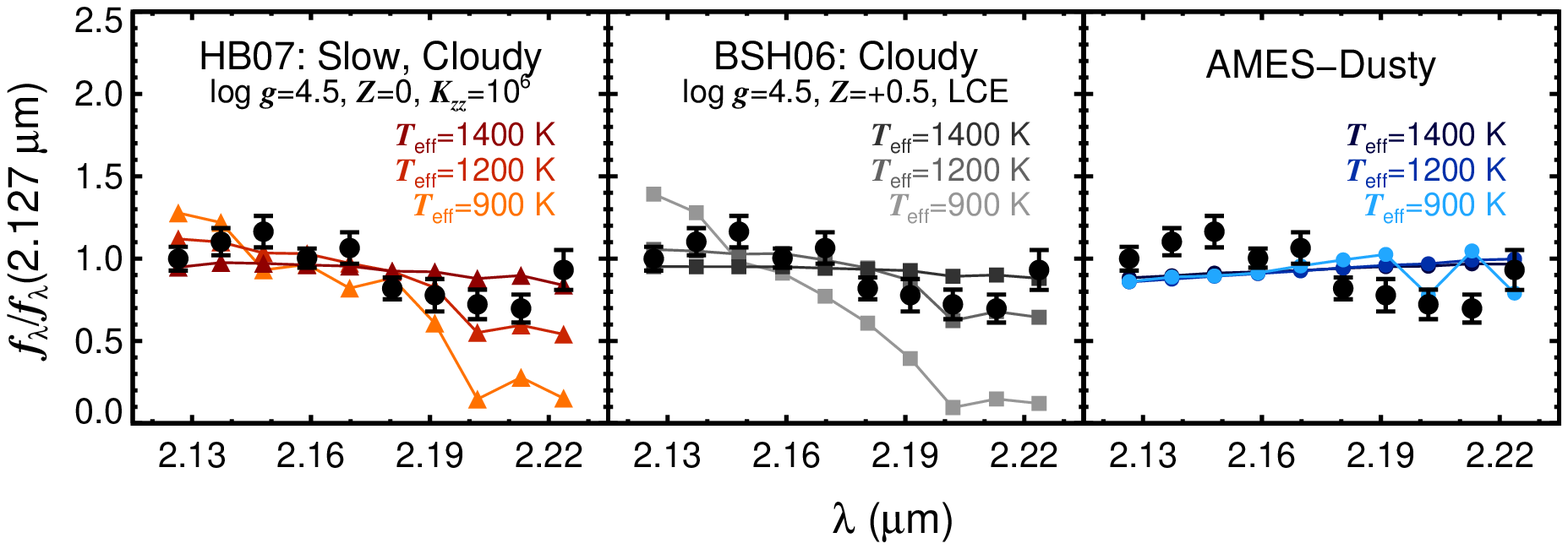}}
\end{center}
  \vskip -1.6 in
  \caption{Examples of atmospheric model fits to our $K$-band spectrum of HR~8799~b (black circles with error bars).   The left panel shows the effect of varying $T_\mathrm{eff}$ in the cloudy version of the $slow$ HB07 models for log~$g$=4.5, $Z$=0, and $K_{zz}$=10$^6$ cm$^2$ s$^{-1}$.  The same temperatures are shown in the middle panel for the cloudy version of the BSH06 models at high metallicities ($Z$=+0.5) and log~$g$=4.5.  Effective temperatures below $\sim$1200~K are inconsistent with our OSIRIS spectrum for the physical parameters sampled in the HB07 and BSH06 model grids.  The right panel shows the same temperatures for the Ames-Dusty models, which are generally featureless in this spectral region.  All models are scaled to the OSIRIS data using Equation~\ref{eq:ck}.  \label{fig:bestfitfig_bur_1} } 
\end{figure*}

In Figure \ref{fig:bestfitfig_bur_1} we show examples of cloudy HB07 $slow$ models with log~$K_{zz}$=6 (left), cloudy BSH06 models with $Z$=+0.5 (middle), and Ames-Dusty models (right) for log~$g$=4.5 and $T_\mathrm{eff}$=900, 1200, and 1400~K.  The 900~K models correspond to effective temperatures derived from evolutionary models (\citealt{Marois:2008p18841}) but produce poor fits to the OSIRIS spectrum.

\begin{figure}
\begin{center}
  \resizebox{3.5in}{!}{\includegraphics{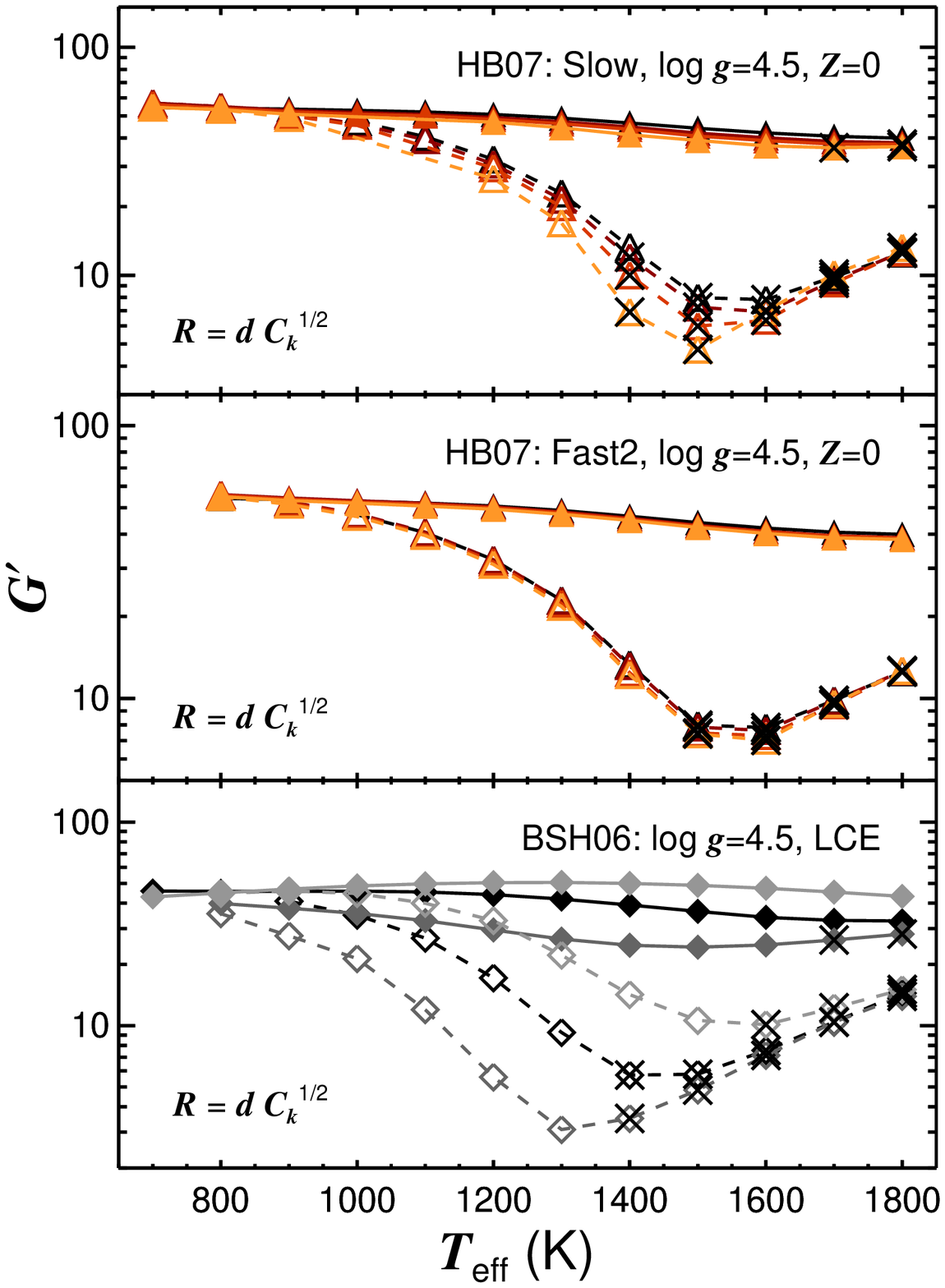}}
\end{center}
  \caption{Atmospheric model fitting to HR~8799~b photometry.  Here the models are scaled by allowing $R$ to vary ($\S$\ref{sec:floatr}).  Models with clear atmospheres are shown with solid lines while those with cloudy atmospheres have dashed lines, and colors are the same as in Figure~\ref{fig:chi2summaryfig}.   Models inconsistent with the upper limits at 3.3~$\mu$m and $M$-band (\citealt{Hinz:2010p20424}) are  marked with crosses.  The best-fitting models have cloudy atmospheres with effective temperatures of 1300-1600~K, high-$K_{zz}$ values (top, middle), and high metallicities (bottom).  \label{fig:sedfits_bur_floatr_summaryfig} } 
\end{figure}

\subsection{Physical Properties from Photometric Analysis}\label{sec:photanal}

To further constrain the physical properties of HR~8799~b, we fit the same three sets of atmospheric models to published photometry.       We use the \citet{Marois:2008p18841} $J$, $H$, $CH_{4}s$, $K_S$, and $L'$ photometry in the fits.  We exclude the published $CH_{4}l$ photometry because the 1.6~$\mu$m CH$_4$ band is too weak in atmospheric models as a result of incomplete CH$_4$ line lists (e.g., \citealt{Saumon:2007p20426}; \citealt{Leggett:2007p68}).  We choose not use other published photometry near $H$-band to avoid overweighting that portion of the spectrum in the fitting ($F160W$: \citealt{Lafreniere:2009p17982}).  As described below, 3~$\sigma$ upper limits at 3.3~$\mu$m and $M$-band by \citet{Hinz:2010p20424} are also incorporated into our analysis.  Our approach is to fit the observed SED using two techniques to scale the predicted flux densities from the atmospheric models. 

\begin{figure}
  \hskip -.7 in
  \resizebox{5in}{!}{\includegraphics{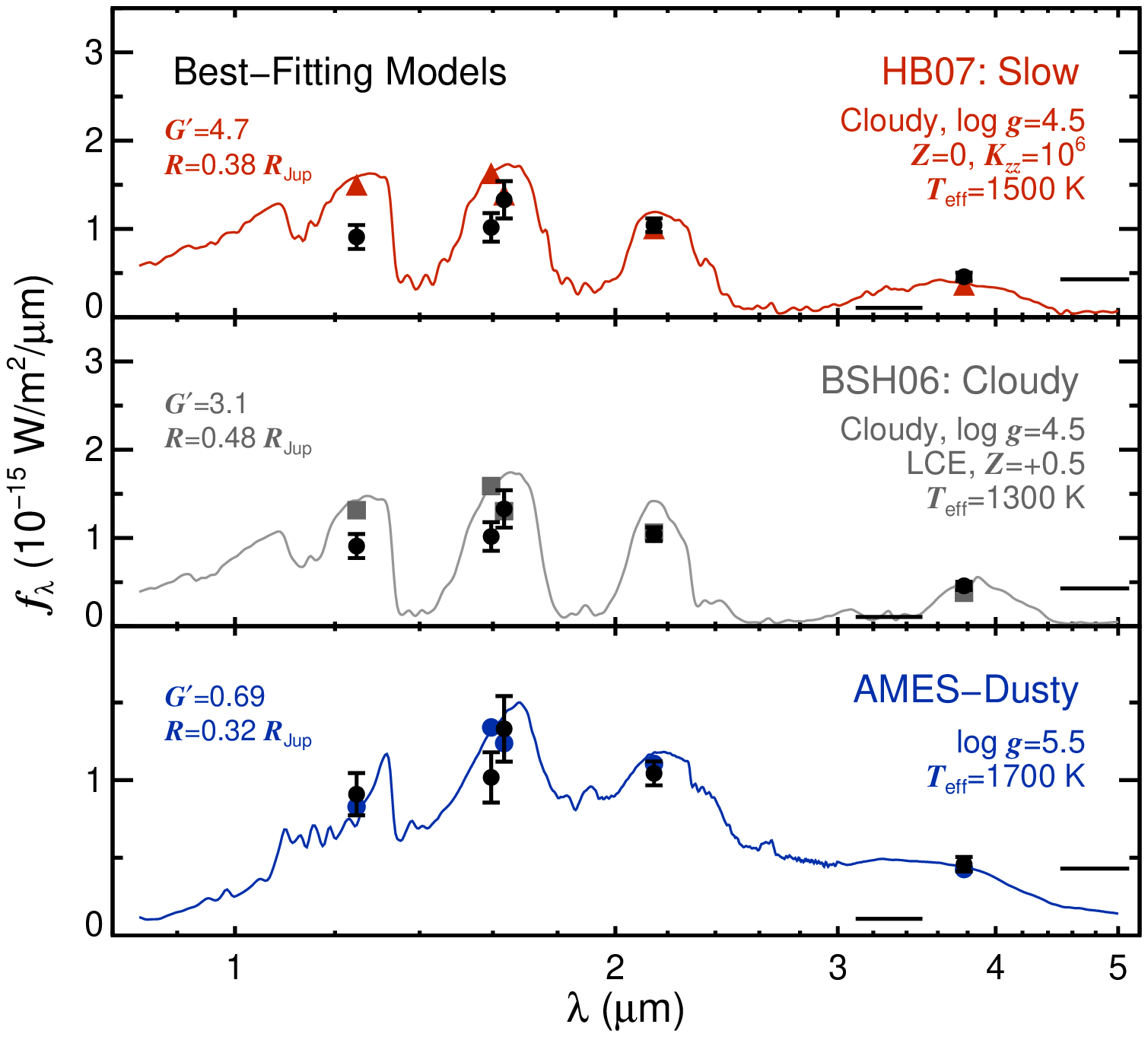}}
  \caption{Best-fitting HB07 $slow$ (top), BSH07 (middle), and Ames-Dusty (bottom) atmospheric models to photometry of HR~8799~b (at 10 pc; black circles with error bars).   The best-fitting models to the detections for the HB07 $slow$ and Ames-Dusty models are inconsistent with the 3~$\sigma$ upper limit at 3.3~$\mu$m from \citet[horizontal lines]{Hinz:2010p20424}.  The effective temperatures and radii inferred from these fits are inconsistent with the evolutionary model-inferred values of 800-900~K and 1.1-1.3~$R_\mathrm{Jup}$.  Although cloudy HB07 and BSH06 models provide better fits than the clear models, the $J$-band flux is still overestimated in the models.  This suggests that even higher dust opacities are required, although not as high as in the Ames-Dusty models, which are inconsistent with the upper limits.  All models have been smoothed for better rendering and the radius is allowed to float in the scaling ($\S$\ref{sec:floatr}). \label{fig:bestfitfig_bur_amesdusty_bestfit} } 
\end{figure}

\subsubsection{Radius as a Free Parameter}\label{sec:floatr}

The first method uses the $G_k^{'}$ statistic (Equation~\ref{eq:gstat}) as a measure of the goodness-of-fit and scales the model values using $C_k^{'}$ from Equation~\ref{eq:gstatck}.  The model spectra represent the emergent flux at the object's surface, so the best-fitting $C_k^{'}$ value is also equal to $(R/d)^2$, where $R$ is the object's radius and $d$ is its distance.  Throughout this work we use HR~8799~b flux densities at 10 pc.  $R$ is therefore allowed to vary in a way that may not satisfy $L$=$4 \pi R^2 \sigma T_\mathrm{eff}^4$.  The model monochromatic flux densities for each filter are computed the same way as in $\S$\ref{sec:bdcomp}.  After we compute goodness-of-fit statistics we then consider the upper limits at 3.3~$\mu$m and $M$-band (\citealt{Hinz:2010p20424}) to infer a range of best-fitting effective temperatures.  The Hinz et al. upper limits are incorporated by considering two extreme cases.  At one extreme, any models with flux above the upper limits are ruled out.  In the weakest scheme, the upper limits are ignored and the best-fitting model is the one that provides the best fit to the detections only.  As shown below, the two extremes produce very similar answers, so we adopt the range of models from both cases as the best fits.   We note that between the 3.3~$\mu$m and $M$-band upper limits, the upper limit at 3.3~$\mu$m provides the stronger constraint on the models and generally disfavors the cloudy versions with effective temperatures between 1400--1800~K.

\begin{figure}
  \hskip -.2 in
  \resizebox{3.5in}{!}{\includegraphics{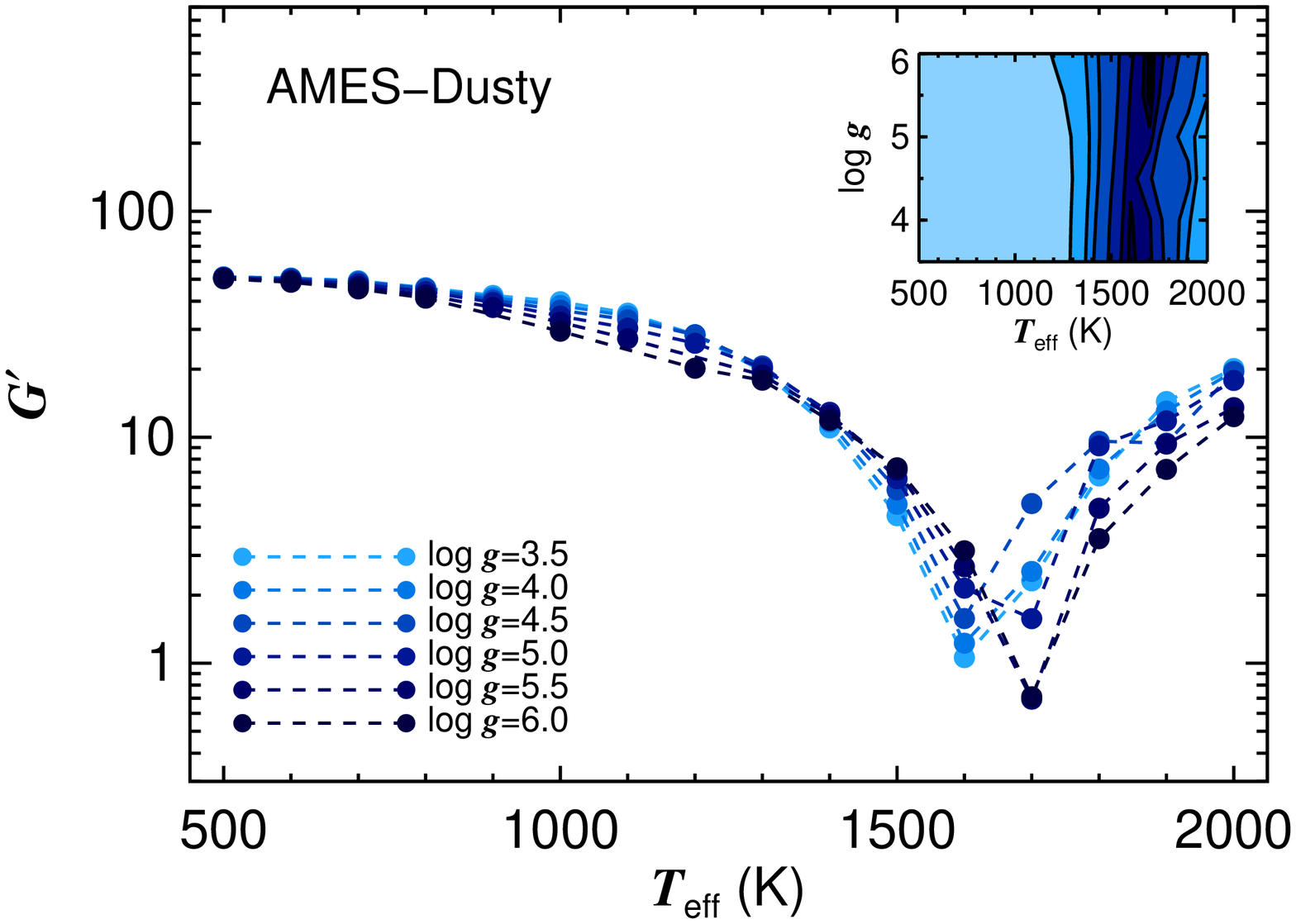}}
  \caption{Ames-Dusty atmospheric model fitting to the photometry of HR~8799~b.  The global minimum is located at log~$g$=5.5 and $T_\mathrm{eff}$=1700~K, and a local minimum exists at log~$g$=3.5 and $T_\mathrm{eff}$=1600~K.  The inset shows the same data as a contour plot for contour levels of \{1.3,2,4,8,15,20,30\} times the minimum $G'$ value (0.69).  Although the best-fitting models fit the detections rather well (indicated by the low $G'$ values), \emph{all models} are inconsistent with the upper limits from \citealt{Hinz:2010p20424}.  The $M$-band upper limit is inconsistent with $T_\mathrm{eff}$$<$900~K while the 3.3~$\mu$m upper limit is inconsistent with models $>$800~K.    The best-fitting model to the detections is shown in Figure~\ref{fig:bestfitfig_bur_amesdusty_bestfit}. \label{fig:sedfits_amesdusty} } 
\end{figure}

Cloudy models produce significantly better fits to the observed SEDs for both the HB07 and BSH06 models (Figure~\ref{fig:sedfits_bur_floatr_summaryfig}).  HB07 models with larger values of $K_{zz}$ yield better fits for a given effective temperature.  The best-fitting models for the HB07 $slow$ grid are the cloudy variants with $T_\mathrm{eff}$=1300-1500~K and log~$K_{zz}$=6, and for the \emph{fast2} grid are the cloudy models with $T_\mathrm{eff}$=1400-1600~K and log~$K_{zz}$=6 (Figure~\ref{fig:bestfitfig_bur_amesdusty_bestfit}).  Based on the scaling factors, the radii inferred for the best-fitting $slow$ models are 0.38-0.50~$R_\mathrm{Jup}$\footnote{The standard value for Jupiter's radius is 71,492~km (\citealt{Lindal:1981p20491}), which is the equatorial radius at 1 bar.} and \emph{fast2} models are 0.35-0.44~$R_\mathrm{Jup}$.  For a given effective temperature, BSH06 models with higher metallicities and clouds result in better fits.  The best-fitting BSH06 model is the cloudy version with $T_\mathrm{eff}$=1300~K and a corresponding radius of 0.48~$R_\mathrm{Jup}$ (in this case the upper limits do not exclude the best-fitting model to the detections only).  The best-fitting Ames-Dusty model to the detections has $T_\mathrm{eff}$=1700~K, log~$g$=5.5, and $R$=0.32~$R_\mathrm{Jup}$ (Figures~\ref{fig:bestfitfig_bur_amesdusty_bestfit} and \ref{fig:sedfits_amesdusty}).  All the Ames-Dusty models, however, do not agree with the upper limits.  The $M$-band upper limit is inconsistent with $T_\mathrm{eff}$$<$900~K and the 3.3~$\mu$m upper limit is inconsistent with models $>$800~K.  Overall, these temperatures and radii are hotter and smaller than those predicted by evolutionary models (800--900~K and 1.1--1.3$~R_\mathrm{Jup}$) given the age and luminosity of the planet (see $\S$\ref{sec:discussion}).\footnote{We also calculate luminosities for each model by integrating the scaled synthetic spectra.  We derive a luminosity log~$L_\mathrm{bol}/L_{\odot}$ of --5.2$\pm$0.1~dex, which is in excellent agreement with the Marois et al. value of --5.1$\pm$0.1~dex.  We note that for lower effective temperatures (larger radii), the luminosities are closer to --5.3~dex, while for higher effective temperatures (smaller radii), the luminosities are near --5.1~dex.}

\begin{figure}
  \hskip -.7 in
  \resizebox{5in}{!}{\includegraphics{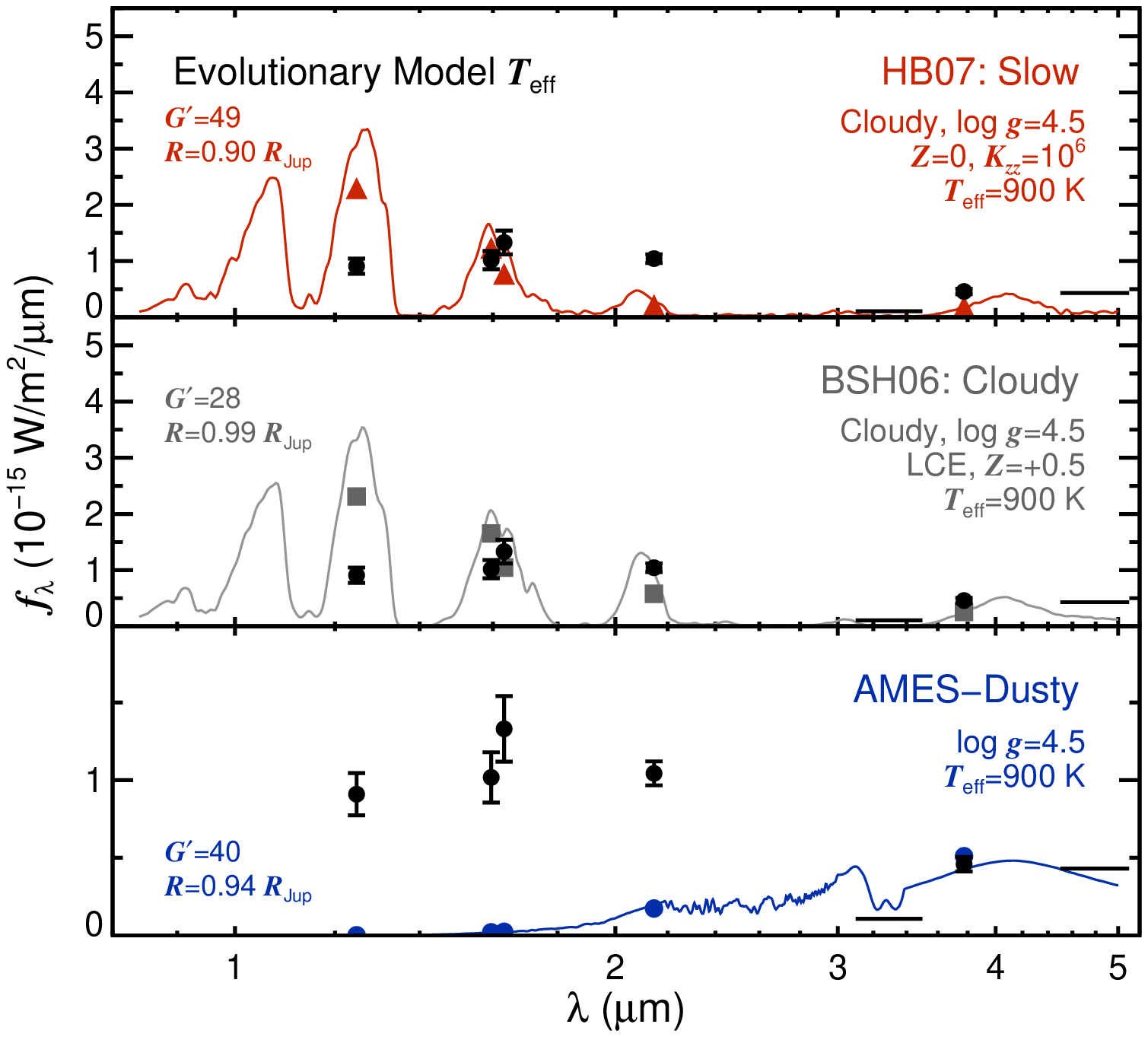}}
  \caption{Comparison of atmospheric models with effective temperatures predicted by evolutionary models ($\sim$900~K) to the photometry of HR~8799~b.  The $J$-band flux is overestimated and the $K$-band flux is underestimated in the HB07 $slow$ (top) and BSH06 (middle) models, and in both cases strong 2.2~$\mu$m CH$_4$ absorption is present, which is not seen in our $K$-band spectrum.  The 900~K Ames-Dusty model provides a very poor fit to the data.  Models with intermediate clouds between the HB07/BSH06 cloudy version and the Ames-Dusty version may provide better fits to the photometry while still inhibiting strong methane absorption and remaining below the 3.3~$\mu$m and $M$-band upper limits.  All models have been smoothed for better rendering and the radius is allowed to float in the scaling ($\S$\ref{sec:floatr}).  \label{fig:bestfitfig_bur_amesdusty_evteff} } 
\end{figure}

In Figure \ref{fig:bestfitfig_bur_amesdusty_evteff} we show the 900~K atmospheric models compared to the HR~8799~photometry, which corresponds to the temperature predicted by the evolutionary models. The $J$-band fluxes are overpredicted and the $K$-band fluxes are underpredicted in the HB07 and BSH06 cloudy models, indicating that even cloudier or more metal-rich models are needed to reproduce the observed flux.  The 900~K Ames-Dusty model is a poor fit to the data, indicating that this limiting case of dust content is too extreme for HR~8799~b.

\begin{figure}
 \resizebox{3.5in}{!}{\includegraphics{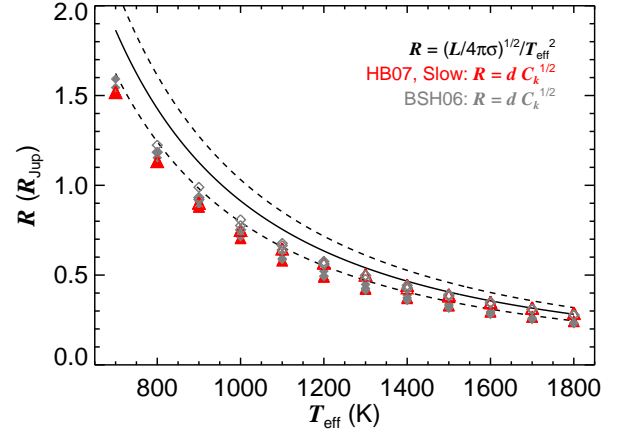}}
  \caption{Radii from fitting atmospheric models to photometry of HR~8799~b.  Values are computed from the scaling factor $C_k^{'}$=($R$/$d$)$^2$ for the HB07 models (red) and BSH06 models (gray).  Cloud-free models are represented with filled symbols and cloudy models with open symbols.  The radii are systematically offset from the values expected based on $L$=4$\pi R^2 \sigma T_\mathrm{eff}^4$ using the luminosity of HR~8799~b (log~$L_\mathrm{bol}/L_{\odot}$=--5.1$\pm$0.1) and the effective temperature of the atmospheric model (black curve and dashed error bar).  The magnitude and significance of the offset for the cloudy models appear to increase at at lower effective temperatures.    \label{fig:radplot} } 
\end{figure}

There appears to be a systematic underestimate of radii inferred using this fitting technique compared to the radius derived using $L$=4$\pi R^2 \sigma T_\mathrm{eff}^4$ (Figure \ref{fig:radplot}).  The magnitude of the offset increases at lower effective temperatures and may result from inadequate atmospheric models.  This suggests that caution must be used when inferring distances or radii from $C_k$ at very low temperatures.

\begin{figure}
  \resizebox{3.5in}{!}{\includegraphics{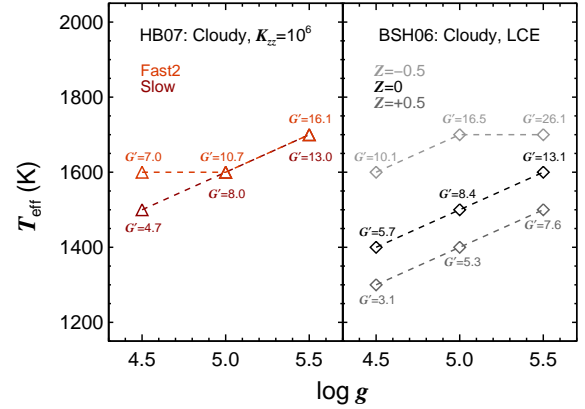}}
  \caption{Best-fitting model atmospheres as a function of the assumed surface gravity.  Each point shows the best-fitting model to the $J$-, $CH_4s$-, $H$-, $K_S$-, and $L'$-band photometry of HR~8799~b. Lower gravities result in better-fitting models (lower $G'$ values) with cooler effective temperatures. The HB07 $slow$ models produce better fits than the \emph{fast2} models (left).  The BSH06 models produce better fits and cooler temperatures at higher metallicities across all gravities (right).  \label{fig:loggfig} } 
\end{figure}

To study the influence of surface gravity on our results, we also fit the HB07 and BSH06 models at log~$g$=5.5 and 5.0 (Figure~\ref{fig:loggfig}).  The distributions of $G'$ values shift to lower effective temperatures at lower gravities.  Additionally, the minima of the distributions occur at lower $G'$ values for lower gravities, indicating that low gravities produce better fits than high gravities.  We note that for the HB07 models the $slow$ models result in lower $G'$ values than \emph{fast2} models for all gravities, and for the BSH06 models the high metallicities are preferred over solar-metallicity and metal-poor models for all gravities.

\subsubsection{Fixed  Radius}\label{sec:fixedr}

Our second fitting method  uses $L$=4$\pi R^2 \sigma T_\mathrm{eff}^4$, the measured luminosity (log~$L_\mathrm{bol}/L_{\odot}$=--5.1~dex), and the model effective temperature to derive the scaling factor for each model.  All models with the same effective temperature will have the same radius for a given luminosity, and therefore the same scaling factor $C_k$=$(R/d)^2$.  To incorporate the uncertainty in the measured luminosity (0.1~dex) we fit the atmospheric models in a Monte Carlo fashion by randomly drawing a new luminosity value, computing a new scaling factor using the model effective temperature, and calculating the $G_k^{'}$ value for each trial.  We use the same weights $w_i$ defined in Equation~\ref{eqn:weights}.  We repeat this for 10$^3$ trails, resulting in a distribution of $G_k^{'}$ values for each model.

The outcome of fitting the models by fixing $R$ at each $T_\mathrm{eff}$ are similar to those from treating $R$ as a free parameter (Figure~\ref{fig:sedfits_bur_ltr_summaryfig}).  Cloudy models, high $K_{zz}$ models, and high metallicity models produce the best fits to the photometry.   However, the systematic differences between the models and the data are similar to those noted in $\S$\ref{sec:floatr}, with the largest discrepancy in the $J$-band.

\section{Discussion}\label{sec:discussion}

Atmospheric model fitting to our OSIRIS spectrum and published photometry suggests that HR~8799~b is an object with extreme physical parameters (see Table~\ref{tab:sum} and Figure~\ref{fig:pictogram} for a summary of our results).  The best-fitting atmospheric models are consistently the cloudy variants, which is in agreement with previous work indicating a high photospheric dust content (\citealt{Marois:2008p18841}; \citealt{Lafreniere:2009p17982}).  The unusually red color of HR~8799~b ($J$--$K_\mathrm{S}$=2.25~mag) is particularly striking given its intrinsic faintness ($M_J$=16.30~mag), which is comparable to mid-T~dwarfs in the field (which have $J$--$K$$<$0~mag; e.g. \citealt{Knapp:2004p15209}).  Its position on the ($M_J$, $J$--$K$) diagram from \citet[their Figure 3]{Burrows:2006p7009} is near the ``Case~A'' and ``Case~B'' model tracks, which formulate clouds in a similarly extreme fashion to the Ames-Dusty models with no gravitational settling of dust.   However, the Ames-Dusty atmospheric models are inconsistent with the upper limits at thermal wavelengths.  The more modest dust model used in HB07 and BSH06 models overestimate the $J$ and $H$ band fluxes and underestimate the $K$ band flux (which are suppressed and enhanced, respectively, in cloudy models), implying that an intermediate case between the cloudy HB07/BSH06 prescription and the Ames-Dusty case will provide better fits to the HR~8799~b photometry.  Other low-mass, low-gravity objects such as HR~8799~d ($M_J$=15.26~mag, $J$--$K_\mathrm{S}$=2.15~mag; \citealt{Marois:2008p18841}) and 2MASS~1207-3932~b ($M_J$=16.38~mag, $J$--$K_\mathrm{S}$=3.07~mag; \citealt{Chauvin:2004p19400}, \citealt{Mohanty:2007p6975}) sit in the same region of the color-magnitude diagram, suggesting that young planetary-mass objects possess radically enhanced cloud properties compared to every class of brown dwarf currently known.

\begin{figure}
\begin{center}
  \resizebox{3.5in}{!}{\includegraphics{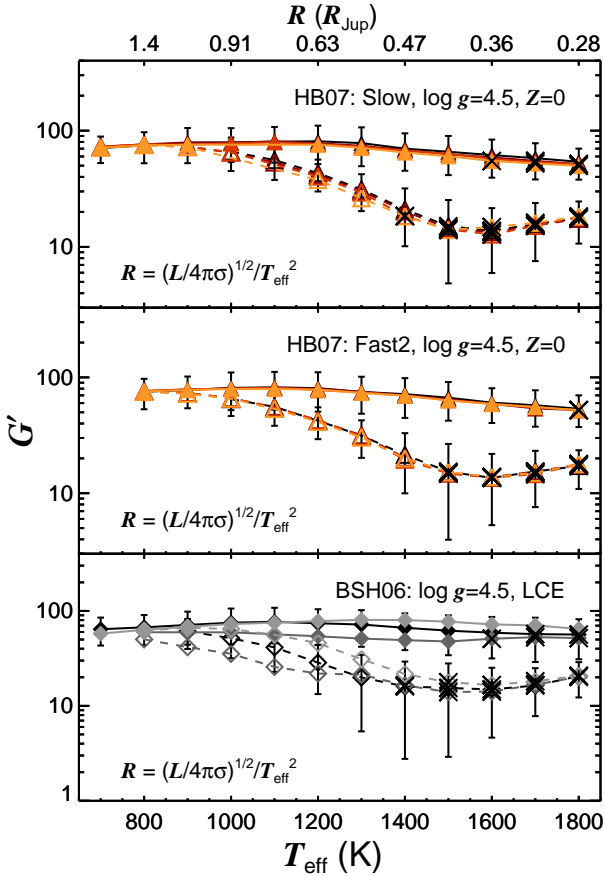}}
\end{center}
  \caption{Best-fitting HB07 and BSH06 atmospheric models when scaled so that the radii are fixed for a given effective temperature (defined using the luminosity of HR~8799~b: log~$L_\mathrm{bol}/L_{\odot}=-5.1\pm$0.1).  Models with the same effective temperatures have the same radii in this fitting scheme.  The results are similar to those when $R$ is allowed to vary freely.  $G'$ values are computed in a Monte Carlo fashion by varying the scaling factor $C'$ based on the luminosity uncertainty ($\S$\ref{sec:fixedr}).  The mean $G'$ values from each distribution are plotted, and for clarity we overplot only the 1~$\sigma$ error bars for the clear and cloudy LCE models.  The color coding is the same as in Figure~\ref{fig:sedfits_bur_floatr_summaryfig}. \label{fig:sedfits_bur_ltr_summaryfig} } 
\end{figure}

Among the cloudy models, we find that those with high $K_{zz}$ values (10$^6$ cm$^2$ s$^{-1}$) and metal-rich compositions ($Z$=+0.5) are generally preferred.  These results are particularly intriguing because field L and T~dwarfs do not appear to have such high levels of disequilibrium chemistry; values of $\sim$10$^2$-10$^6$~cm$^2$~s$^{-1}$ typically produce the best fits to near- and mid-infrared photometry and spectroscopy of ultracool dwarfs (\citealt{Leggett:2007p18996}; \citealt{Saumon:2007p20426}; \citealt{Stephens:2009p19484}; \citealt{Geballe:2009p17672}).  The gas giants in our solar system have moderate values of $K_{zz}$ in their stratospheres ($\sim$10$^2$-10$^5$ cm$^2$ s$^{-1}$;  \citealt{Saumon:2006p10455})\footnote{Note that $K_{zz}$ is much higher in the convective regions of solar system giant planetary atmospheres than the radiative zone, reaching values of $\sim$10$^6$-10$^9$ cm$^2$ s$^{-1}$ (\citealt{Fegley:1994p20765}; \citealt{Atreya:1999p20430}; \citealt{Moses:2005p20496}; \citealt{Saumon:2007p20426}).  $K_{zz}$ also varies as a function of latitude and atmospheric pressure; see \citet{Moses:2005p20496} for a detailed discussion applied to Jupiter's atmosphere.} with substantially metal-rich atmospheric compositions  (metal abundances range from $\sim$2-4 times solar values for Jupiter, up to $\sim$10 times for Saturn, and even higher for some species in Uranus and Neptune; \citealt{Atreya:2003p20507}; \citealt{Lodders:2004p20281}; \citealt{Flasar:2005p20509}; \citealt{Fortney:2009p20502}).

\begin{figure}
\begin{center}
  \hskip -0.4 in
  \resizebox{3.5in}{!}{\includegraphics{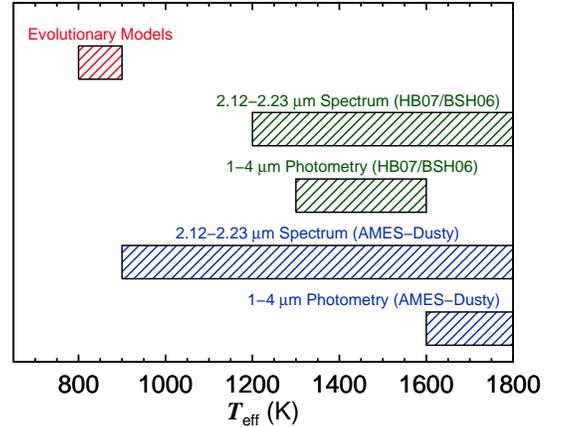}}
\end{center}
  \caption{Graphical representation of our atmospheric model fitting results.  The HB07/BSH06 model fits (green) to our OSIRIS spectrum imply effective temperatures $\gtrsim$1200~K, while fits to the 1-4~$\mu$m photometry suggest effective temperatures between 1300~K and 1600~K.  Ames-Dusty model fits (blue) to the spectrum provide poor constraints on the effective temperature ($\gtrsim$800~K).  The best-fitting Ames-Dusty model to the 1-4~$\mu$m photometric detections of HR~8799~b is $T_\mathrm{eff}$=1700, log~$g$=5.5, although all models in this grid are inconsistent with the upper limits at 3.3~$\mu$m and $M$~band from \citet{Hinz:2010p20424}.  The evolutionary models of \citet[red]{Baraffe:2003p588} predict temperatures of 800-900~K.      \label{fig:pictogram} } 
\end{figure}

\begin{deluxetable}{lc}
\tablewidth{0pt}
\tablecolumns{2}
\tablecaption{Summary of Results \label{tab:sum}}
\tablehead{
                 \colhead{Type of Fit}   &    \colhead{Best Fits}        }   
\startdata
\cutinhead{$K$-band Spectroscopic Analysis}
Field Brown Dwarfs     &     SpT=T2 (earlier than T4)       \\
HB07 Models                &       $T_\mathrm{eff}$$>$1100 K \\
BSH06 Models                &       $T_\mathrm{eff}$$>$1100 K  \\
Ames-Dusty Models      &      $T_\mathrm{eff}$$>$800 K  \\
\cutinhead{1-4 $\mu$m SED Analysis}
Field Brown Dwarfs     &    SpT between $\sim$L5-L8       \\
HB07 Models                &       $T_\mathrm{eff}$$\sim$1300-1600 K, cloudy, high $K_{zz}$ \\
BSH06 Models                &       $T_\mathrm{eff}$$\sim$1300 K, cloudy, $Z$=+0.5 \\
Ames-Dusty Models     &       $T_\mathrm{eff}$=1700~K, log~$g$=5.5\tablenotemark{a} \\
\enddata
\tablenotetext{a}{This is the best-fitting model to the detections, but all Ames-Dusty models are inconsistent with the upper limits at  3.3~$\mu$m and $M$ band from \citet{Hinz:2010p20424}.}
\end{deluxetable}

The metallicity of HR~8799~b has implications for scenarios of its formation.  Giant planets formed through core accretion (\citealt{Pollack:1996p19730}; \citealt{Alibert:2005p17987}) are generally predicted to have metal-rich atmospheres as a result of planetesimal accretion, the accretion of metal-rich gas, and/or core erosion (\citealt{Alibert:2005p20482}; \citealt{Guillot:2006p20515}; \citealt{Mousis:2009p20511}; see \citealt{Fortney:2008p8729} for a thorough summary).  There is currently no theoretical consensus regarding the atmospheric metallicities of gas giants formed through  disk instability, the proposed alternative mechanism of giant planet formation (\citealt{Cameron:1978p20284}; \citealt{Boss:1997p18822}), but recent simulations by \citet{Helled:2010p20510} suggest that the metallicity of the HR~8799 planets should be similar to HR~8799 itself if they formed in this fashion.  Unfortunately, the metallicity of HR~8799 is not known.  HR~8799 is a $\lambda$~Bootis star, so while its photosphere is metal-poor ([M/H]$\sim$--0.5; \citealt{Gray:1999p20164}; \citealt{Sadakane:2006p20501}), this may be a result of recent accretion of metal-poor gas and its internal metallicity could be very different (see \citealt{Moya:2010p20492} for a detailed discussion of this phenomenon).  Ultimately, more refined predictions from planet formation theory combined with a better determination of the internal metallicity of HR~8799 and the metallicities of its planets will provide a key test to distinguish between core accretion and disk instability as the formation mechanism.\footnote{Note that the luminosities of HR~8799~b, c, and d do not agree with the luminosities of giant planets formed in the ``cold start'' models of \citet{Marley:2007p18269} and \citet{Fortney:2008p8729} for any ages or planet masses.  The planets are far too luminous compared to the models.  The agreement with the ``hot-start'' models may suggest that the HR~8799 planets were formed through disk instability, as has been suggested by \citet{DodsonRobinson:2009p19734} based on an analysis of plausible theoretical formation mechanisms. }

While our model fitting of the SED demands a very cloudy atmosphere, the other physical parameters are less well-constrained.  We caution that it is not clear whether our results represent actual evidence for strong vertical mixing and high metallicity, or whether they are simply an outcome of using grids of models with only a limited variety of cloud prescriptions (cloud-free, 100 $\mu$m forsterite clouds, and completely cloudy).   No atmospheric model that we analyzed agrees well with the SED of HR~8799~b, indicating that the atmospheric models are incorrect or that the physical parameters of HR~8799~b fall outside of the available grids.  The relative success of these atmospheric models in reproducing the spectra of field brown dwarfs (e.g., \citealt{Cushing:2008p2613}) suggests that the reason for the disagreement between atmospheric models and the photometry is most likely caused by a higher dust opacity in HR~8799~b than is included in the models.  Indeed, there are many parameters in the cloud models that can influence the emergent spectra, such as particle size and cloud model prescription (see, e.g., \citealt{Burrows:2006p7009} and \citealt{Helling:2008p14061} for detailed discussions), and thus the best-fitting vertical mixing and/or metallicity value may change with a broader set of models.

Our  results highlight a significant discrepancy between atmospheric and evolutionary model predictions.  The measured luminosity of planet~b (log~$L_\mathrm{bol}/L_{\odot}$=--5.1$\pm$0.1 dex) combined with the age of HR~8799 ($\sim$30-160~Myr) implies an effective temperature of $\sim$800-900~K,\footnote{Note that the effective temperatures predicted by evolutionary models based on the luminosity of HR~8799~b are relatively insensitive to age since the radii of brown dwarfs and giant planets are nearly constant with mass and age after $\sim$100~Myr.  At 1~Gyr the predicted temperature is $\sim$1000~K and at 10~Gyr it is $\sim$1100~K (\citealt{Baraffe:2003p588})}  a radius of $\sim$1.16-1.24~$R_\mathrm{Jup}$, and a mass of $\sim$7-10 $M_\mathrm{Jup}$ based on the \citet{Baraffe:2003p588} evolutionary models.  However, we find the best-fitting atmospheric models have effective temperatures between 1300-1700~K and radii between $\sim$0.3-0.5~$R_\mathrm{Jup}$, significantly hotter and smaller than the evolutionary model predictions.  This discrepancy for HR~8799~b has previously been noted by \citet{Marois:2008p18841}, \citet{Lafreniere:2009p17982}, and \citet{Hinz:2010p20424}.  In contrast, the radii inferred from fitting the spectra of late-M dwarfs (which have similar radii of $\sim$1~$R_\mathrm{Jup}$) have been in better agreement with the values predicted by evolutionary models (\citealt{Leggett:2000p19460}; \citealt{Bowler:2009p19621}; \citealt{Dupuy:2009p19489}) and likewise for field L and T dwarfs (\citealt{Stephens:2009p19484}).

The radii inferred from atmospheric model fitting of HR~8799~b are likely to be incorrect because giant planets with degenerate cores cannot reach such small volumes at any point in their lifetimes (e.g., \citealt{Guillot:2005p20459}; \citealt{Fortney:2007p19757}).  Moreover, evolutionary models are less sensitive to boundary conditions (atmospheric parameters) than synthetic spectra are (\citealt{Chabrier:2000p161}; \citealt{Saumon:2008p14070}), which implicates the atmospheric models as the culprit.  One probable explanation is that HR~8799~b has atmospheric properties that are outside of the parameter space explored by current models.  At cool temperatures $\lesssim$1000~K (large radii), the 100 $\mu$m cloud prescription from BSH06 does not significantly improve the fit to the photometry compared to cloud-free models (Figures  \ref{fig:sedfits_bur_floatr_summaryfig} and \ref{fig:sedfits_bur_ltr_summaryfig}).  However, increasing the metallicity not only improves the fit at those temperatures, but it also shifts the best-fitting model from higher temperatures to lower temperatures.  Even higher metallicities and dustier atmospheres (but less extreme than the Ames-Dusty case) may yield a best-fitting model with a temperature and radius closer to the values predicted by evolutionary models and with methane still inhibited in the $K$-band spectrum.

An alternative solution to the discrepancy between the atmospheric model fitting results and predictions from evolutionary models is that HR~8799~b is underluminous, perhaps as a result of obstruction by an edge on-disk as suggested by \citet{Marois:2008p18841}.  This scenario is unlikely, however, because a similar underluminosity is seen with HR~8799~c and d.  A radius discrepancy has also been found with 2MASS~1207-3932~b, which \citet{Mohanty:2007p6975} and \citet{Patience:2010p20422} interpret as an underluminosity caused by an edge-on disk.  However, in light of the analogous problems with the HR~8799 planets, we favor a common origin either from fitting imperfect atmospheric models, or fitting models that do not cover sufficient parameter space.  No underluminosity is observed for the brown dwarf primary 2MASS~1207-3932~A, which also makes this edge-on disk scenario unfavorable.  More extensive atmospheric model development and fitting to the HR~8799 planets and other low-mass, low-gravity objects using different cloud prescriptions, metallicities, and eddy diffusion coefficients may help resolve this issue.

Our spectroscopic and photometric analysis indicates that the spectral type for HR~8799~b falls in the late-L/early-T range, corresponding to a physical regime where the atmospheres of field dwarfs are progressing from cloud-filled to cloudless (\citealt{Burgasser:2002p4157}). The fainter absolute magnitude of HR~8799~b compared to field objects appears to be an extension of the suggestion by \citet{Metchev:2006p10342} that the L/T transition depends on surface gravity (or equivalently age). In their study of HD~203030B, a companion to a young ($\sim$130-400~Myr) field star, they found a temperature of $\sim$100-200 K cooler for this L/T transition object compared to field objects of comparable spectral type using temperatures derived from evolutionary models combined with the age estimate of HD~203030.  The temperatures inferred for the HR~8799 planets from evolutionary models are even cooler compared to field objects (by 300--400~K). Precise temperature derivations for brown dwarf binaries with dynamical mass measurements support this idea of a gravity-dependent L/T transition (\citealt{Liu:2008p14548}; \citealt{Dupuy:2009p18533}). \citet{Stephens:2009p19484} find additional evidence based on atmospheric model fits to field L and T dwarfs, and \citet{Saumon:2008p14070} make a similar suggestion based on the location of Pleiades L/T transition objects in color-magnitude diagrams.  We note that the same result is obtained with 2MASS~1207-3932~b; its luminosity (log~$L_\mathrm{bol}/L_{\odot}$=--4.72$\pm$0.14;  \citealt{Mohanty:2007p6975}) combined with the likely age of 5-10~Myrs (constrained through membership with the TW Hydrae association) implies an effective temperature of $\sim$1000~K based on evolutionary models (\citealt{Baraffe:2003p588}).  This temperature is well below the L/T transition temperature for field objects ($\sim$1200-1400, \citealt{Golimowski:2004p15703}; \citealt{Cushing:2008p2613}), yet the spectral type of 2MASS~1207-3932~b is mid- to late-L (\citealt{Chauvin:2004p19400}; \citealt{Mohanty:2007p6975}).

\section{Summary}\label{sec:summary}

We obtained a 2.12-2.23 $\mu$m spectrum of HR~8799~b with the OSIRIS adaptive optics integral field spectrograph at Keck II.  We performed an empirical comparison of field L and T dwarfs with our OSIRIS spectrum and previously published 1-4~$\mu$m photometry using the SpeX Prism Spectral Library and the compilation of ultracool dwarf photometry from \citet{Leggett:2010p20094}.  Additionally, we fit these data using the HB07 model atmospheres, which include the effects of non-equilibrium chemistry caused by vertical mixing; the BSH06 models, which include non-solar metallicity chemical abundances; and the Ames-Dusty models, which explore the limiting case of extreme photospheric dust content.  Our results are summarized below, in Table~\ref{tab:sum}, and in Figure~\ref{fig:pictogram}.

$\bullet$ We rule out the presence of strong methane absorption at 2.2~$\mu$m in HR~8799~b.  The best-fitting spectral type to our OSIRIS spectrum is T2, although types earlier than T4 are consistent with our data.

$\bullet$ Empirical fits of the SpeX Prism Library to the $J$-, $CH_4s$-, $H$-, and $K_S$-band photometry of HR~8799~b suggest a spectral type of L5-L8.  The best-fitting field objects are red L dwarfs, which generally show evidence for low gravities and/or unusually dusty atmospheres.

$\bullet$ Empirical fits of the \citet{Leggett:2010p20094} compilation of ultracool dwarf photometry to HR~8799~b $J$-, $H$-, $K_S$-, and $L'$-band photometry suggest a spectral type of mid-L to late-L.  Late-type, red L dwarfs provide the best empirical fits to the near- to mid-infrared SED of HR~8799~b.  Similarly, HR~8799~b lies near red L dwarfs in the $J$--$H$/$K_S$--$L'$ diagram, although it has even redder $J$--$H$ colors than typical red L dwarfs.

$\bullet$ Atmospheric model fits to our $K$-band spectrum imply effective temperatures $>$1100~K.

$\bullet$ Atmospheric model fits to the $J$, $CH_4s$, $H$, $K_S$, and $L'$ bands favor cloudy HB07/BSH06 models with effective temperatures of 1300-1600~K and radii of 0.3-0.5~$R_\mathrm{Jup}$, although no models agree well with all the data.  The best-fitting Ames-Dusty model has $T_\mathrm{eff}$=1700~K, log~$g$=5.5, and $R$=0.32~$R_\mathrm{Jup}$, but all models in this grid are inconsistent with the published upper limits at either 3.3~$\mu$m or $M$ band.  These effective temperatures and radii are inconsistent with evolutionary model predictions of 800-900~K and 1.1-1.3~$R_\mathrm{Jup}$ based on the luminosity of HR~8799~b and the likely age of HR~8799.  The origin of the discrepancy is unclear, but we suggest it is likely a result of imperfect atmospheric models or inadequate range of physical parameters in the models.

$\bullet$ HB07/BSH06 atmospheric models with lower surface gravities, higher $K_{zz}$ values, and higher metallicities generally provide better fits to the photometry of HR~8799~b.  The effective temperature of the best-fitting model decreases as the surface gravity is lowered and as the metallicity is increased.

$\bullet$ Altogether, we find that HR~8799~b has a spectral type consistent with L5-T2.  With an evolutionary model-derived temperature of 800-900~K, HR~8799~b provides further evidence that the L/T transition is a gravity-dependent (or, equivalently, age-dependent) phenomenon.

Future photometric and spectroscopic observations can help to further constrain the physical properties of HR~8799~b.  Cloudy atmospheric models  indicate that the $Y$-band flux should be suppressed in a similar fashion as the observed $J$-band flux.  The $M$-band region is sensitive to non-LCE chemistry; unfortunately, none of the HR~8799 planets were detected in a 2.7~hr observation in $M$-band by \citet{Hinz:2010p20424}.     The lack of strong CH$_4$ absorption at 2.2~$\mu$m in our observations suggests that the CO feature of HR~8799 should be strong.  Spectroscopy targeting the $H$-band methane and the 2.3~$\mu$m CO feature can provide even further constraints on the effective temperature and nonequilibrium chemistry of HR~8799~b.  These observations, combined with spectroscopy of HR~8799~c and d, will elucidate the physical properties of this emerging class of low-mass objects which are characterized by low surface gravities,  low luminosities, and exceptionally cloudy atmospheres.

 \acknowledgments

We thank the referee for their thorough analysis and helpful comments, as well as Beth Biller and Adam Kraus for helpful discussions.  We are grateful to Adam Burrows, Ivan Hubeny, and David Sudarsky for the distributing their atmospheric models to the public.  Additionally, we thank James Larkin, Shelly Wright, and the OSIRIS team for creating and maintaining the OSIRIS data reduction pipeline.  It is a pleasure to acknowledge Al Conrad, Jim Lyke, Jason McIlroy, and the Keck Observatory staff for assistance with our observations.  Katelyn Allers, Jenny Patience, Micka\"{e}l Bonnefoy, and David Lafreni\`{e}re kindly provided us with their published spectra of young L-type objects.   Our research has made use of the SIMBAD database, operated at CDS, Strasbourg, France.  BPB, MCL, and TJD acknowledge support from NSF grants AST-0507833 and AST09-09222.  The authors wish to recognize and acknowledge the very significant cultural role and reverence that the summit of Mauna Kea has always had within the indigenous Hawaiian community.  We are most fortunate to have the opportunity to conduct observations from this mountain.  

\newpage

\appendix

\section{$K_\mathrm{MKO}$ to $K_S$ Conversion}\label{app:kmkotoks}

\begin{figure}
  \begin{center}
  \resizebox{3.5in}{!}{\includegraphics{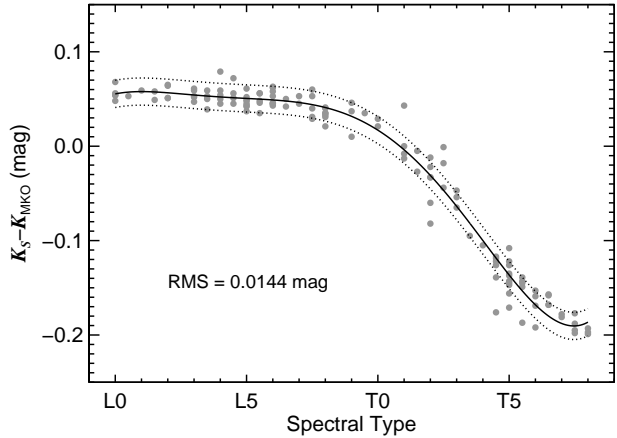}}
  \end{center}
  \caption{$K_\mathrm{MKO}$--$K_S$ as a function of spectral type.  Differential magnitudes are computed from synthetic photometry of objects in the SpeX Prism Spectral Library.  The black curve shows the best-fitting fifth order polynomial to the data, and the rms scatter (0.0144~mag) is shown as dotted curves. \label{fig:ktoksfig} } 
\end{figure}

We converted $K_\mathrm{MKO}$-band to $K_S$-band magnitudes for our sample of L and T dwarfs from \citet{Leggett:2010p20094} by performing  synthetic photometry to objects in the SpeX Prism Spectral Library and fitting a fifth order polynomial to the differential magnitudes (Figure~\ref{fig:ktoksfig}).  Optical types are used for L dwarfs when available, otherwise near-infrared spectral types are used.  The best-fitting fifth-order polynomial is given by $K_S$--$K_\mathrm{MKO}$=$\sum_{i=1}^6 a_i SpT^{i-1}$, where $a$=\{0.05538, 0.0055409, $-0.0038657$, 0.00083746, $-7.5248$$\times$10$^{-5}$, 2.0778$\times$10$^{-6}$\} and $SpT$ is the numerical spectral type defined such that $SpT$=0 for L0 (e.g., L7=7, T0=10).  This relation is valid from L0 to T7, and the rms of the fit is 0.0144 mag.

\section{Fitting Models to Data With Errors in Both Quantities}\label{app:gstatckdp}

When there are uncertainties in both the model and the data we use the goodness-of-fit statistic $G''$ (Equation~\ref{eq:gstat2}), which scales both the model and the model errors by a scaling factor $C_k^{''}$.  Similar to $C_k$ (Equation~\ref{eq:ck}) and $C_k^{'}$ (Equation~\ref{eq:gstatck}), $C_k^{''}$ is computed by minimizing the $G''$ merit function.  Taking the derivative of $G''$ with respect to $C_k^{''}$, equating the result to zero, and simplifying, we get

\begin{equation}\label{eq:gstatckdp}
\sum_{i=1}^{n} w_i \frac{    f_i   \mathcal{F}_{k,i}    \sigma_{f,i}^2   + C_k^{''} ( \sigma_{ \mathcal{F}_{k,i} }^2 f_i^2 -  \sigma_{f,i}^2   \mathcal{F}_{k,i}^2)     -   (C_k^{''})^2 f_i  \mathcal{F}_{k,i}   \sigma_{ \mathcal{F}_{k,i} }^2              }{(  \sigma_{f,i}^2 + ( C_k^{''} \sigma_{ \mathcal{F}_{k,i} } )^2        )^2} = 0,
\end{equation}

\noindent where  $f_i$ is the model flux density for filter $i$ (of $n$ filters), $\mathcal{F}_{k,i}$ is the flux density for object $k$ in filter $i$, and $\sigma_{f,i}$ and $\sigma_{ \mathcal{F}_{k,i} }$ are the uncertainties in both quantities.

We solve for $C_k^{''}$ in an iterative fashion for each object with weights defined by Equation~\ref{eqn:weights}.  $C_k^{''}$ was usually within 30\% of $C_k^{'}$ when $C_k^{'}$ was computed with the ``model'' photometric errors ignored.  Note that when $\sigma_{ \mathcal{F}_{k,i}}$=0, $C_k^{''}$ becomes $C_k^{'}$, and additionally when $w_i$=1, $C_k^{''}$ becomes $C_k$.

\newpage


\end{document}